\documentclass[epj]{svjour}
\usepackage{graphics}
\usepackage{color}
\usepackage{lineno}
\usepackage{cite}
\begin{document}
\title{Systematic studies and optimization of super sensitivity gaseous detectors of sparks, open flames and smoke }
\author{Marcello Abbrescia\inst{1}, Giacinto De Cataldo\inst{2,3},  Antonio di Mauro\inst{2}, Paolo Martinengo\inst{2}, Cosimo Pastore\inst{3}, Vladimir Peskov\inst{2,4}, Francesco Pietropaolo\inst{2,5}, Giacomo Volpe\inst{1} \and Igor Rodionov\inst{4,6}
}                     
\offprints{}          
\institute{Dipartimento Interateneo di Fisica ``M. Merlin'' and Sezione INFN, Bari, Italy \and European Organization for Nuclear Research (CERN), Geneva, Switzerland \and INFN, Sezione di Bari, Bari, Italy \and Institue for Chemical-Physics, Moscow, Russia \and INFN, Sezione di Padova, Padua, Italy \and REMOS Electronic company, Dubna, Russia}
\date{Received: date / Revised version: date}
%
\abstract{
A review of the current progress in developments and tests of supersensitive gaseous detectors of open flame, sparks and smoke is given. A focus on the latest developments is on flat panel type sensors. This design, after further modification, will allow building not only a high efficiency  detector but also offers UV imaging capability for flame visualization and several other applications. These studies were done in the framework of ATTRACT-SMART project.%
\PACS{
      {PACS-key}{discribing text of that key}   
     } 
} 
\maketitle
\section{Introduction}
\label{intro}

In a fight against fire, it is very important to detect its appearance on the earliest possible stage to effectively implement all necessary measures for its suppression.
There are various commercial flame detectors on the market, sensitive either to the flame radiation or to the increase of a local air temperature.
Detectors, sensitive to radiation are more expensive than the thermal one, however their main advantage is that they can record appearance of a fire remotely and on a relatively large distance. There are several types of such detectors sensitive either to ultraviolet (shorter than 280 nm) or to near infrared radiation (0.7$-$1.1 $\mu$m) or to infrared (1.1 $\mu$m and higher), or their combination. These spectral intervals have great advantages compared to visible spectra due to the better flame recognition on strong sun light background. These flame detectors can respond faster and more accurately than a smoke or a heat detector due to the mechanisms it uses to record the flame. 

Ultraviolet (UV) sensitive flame detectors operates in the spectral interval range 185$-$250 nm, where the Sun radiation is fully blocked by the ozone in the upper layer of the atmosphere, however on the ground level air is quite transparent, allowing efficient detection of the UV light emitted by flames. UV sensors are classified by their sensitivity. The highest sensitivity in EU standards have so-called ``class 1'' detectors: they should be capable to detect a $\approx$ 30$\times$30$\times$30~cm$^3$ flame on a distance of 20 m during 20 sec. An example of the Class1 flame sensor is widely used Hamamatsu UVtron R2868.
It is a gaseous detector with metallic cathode filled with pure Ar. UV photons, emitted by a flame, via a photo-effect extract electrons from the cathode and they trigger a glow discharge in the detector, which is quenched by an externally connected resistor. The voltage drop on this resistor generate a signal, the amplitude of which is constant and independent on the number of primary photoelectrons, triggering the glow discharge. Therefore, this detector is operating in a digital mode and consequently cannot distinguish between the single photons, many photons or cosmic rays. The dead time of such a sensor is about 1 ms. 

Coming from the experience gained during developments of photosensitive gaseous detectors for the LAA project \cite{peskov2007new} and for the ALICE RICH detector \cite{Hoedlmoser:2006ra}, a new concept of a flame sensor was suggested: a proportional gaseous detector, operating in quenched gases (mixtures of molecular and noble gas) and combined with a special high sensitivity UV photocathode \cite{Carlson:2002zz}.

\begin{table}[]
\centering
\caption{Counting rate produced by the flame of a candle as a function of distance for three detectors: Hamamatsu R2868, our industrial prototype (sealed detector) and our laboratory prototype (gas flushed detector).}
\begin{tabular}{|l|l|l|l|l|l|}
\hline
\multicolumn{2}{|c|}{\textbf{Hamamatsu R2868}}                                            & \multicolumn{2}{c|}{\textbf{Our industrial prototype}}                                    & \multicolumn{2}{c|}{\textbf{Our lab. prototype}}                                          \\ \hline
Distance (m) & \begin{tabular}[c]{@{}l@{}}Mean number of \\ counts in 10 sec\end{tabular} & Distance (m) & \begin{tabular}[c]{@{}l@{}}Mean number of\\  counts in 10 sec\end{tabular} & Distance (m) & \begin{tabular}[c]{@{}l@{}}Mean number of\\  counts in 10 sec\end{tabular} \\ \hline
1.0          &                                                                            &     1.0         &                             81579                                      &              &                                                                            \\ \hline
1.1          &             583                                                         &                 &                                                                            &                 &                                                                            \\ \hline
2.5          &              99                                                           &                 &                                                                            &                &                                                                            \\ \hline
3.0          &              76                                                           &       3.0       &                             9015                                        &     3.0        &                    87574                                                        \\ \hline
4.5          &               28                                                          &                  &                                                                            &                   &                                                                            \\ \hline
10.0         &               6                                                        &       10.0       &                           811                                            &      10.0        &               7902                                                             \\ \hline
20.0         &                                                                            &                     &                                                                           &                  &                                                                            \\ \hline
30.0         &                                                                            &     30.0         &                           92                                           &       30.0       &      876                                                                      \\ \hline
\end{tabular}
\label{table1}
\end{table}

One of the first tested option was just a single-wire counter with CsI photocathode operating in gas flushed mode. This prototype exhibited extraordinary characteristics, for
example both its sensitivity and the time resolution was 1000 times higher than that obtained by the UVtron R2868, encouraging further developments in this direction. Several other gas flushed prototyped were tested later on.
However, from the practical point of view sealed detectors are needed. In this paper we report our latest progress in this direction.

\section{Sealed single-wire flame detectors}
\label{sec:1}
\subsection{Detectors with CsI or CsTe photocathodes}
\label{sec:2.1}

\begin{figure*}
\centering
\includegraphics{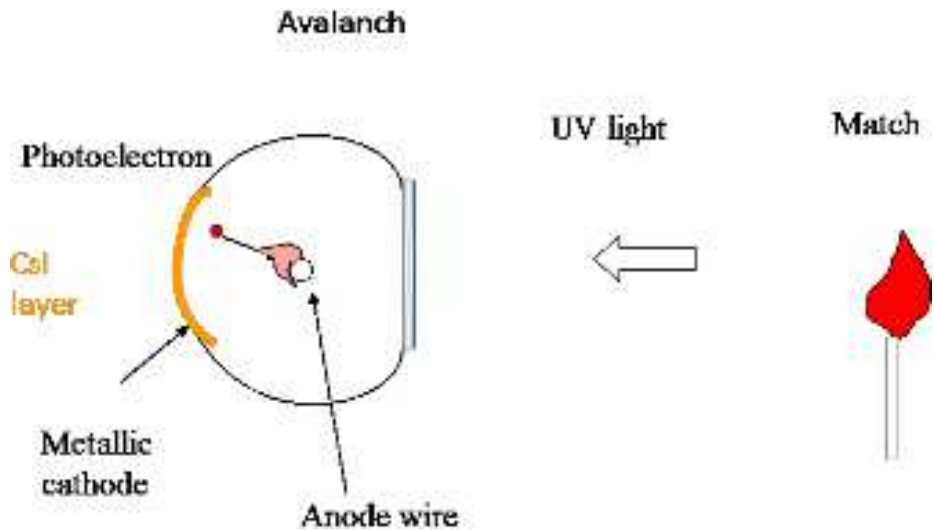}
\caption{Schematics of a single-wire counter with a CsI photocathode}
\label{WorkingPrinciple}       
\end{figure*}

\begin{figure*}
\centering
\resizebox{0.7\textwidth}{!}{
\includegraphics{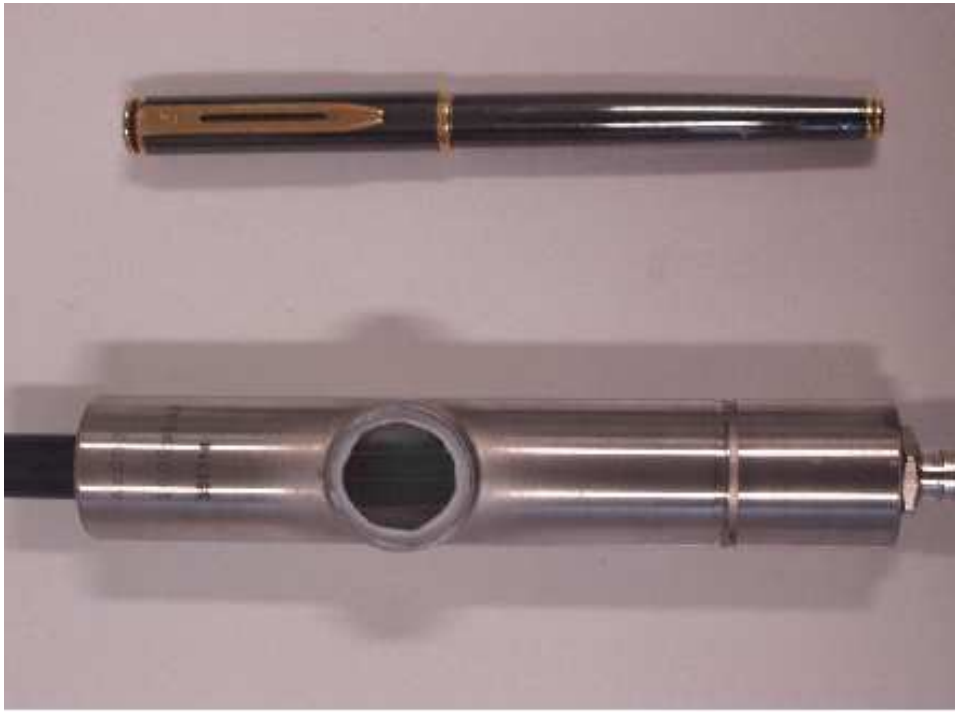}}
\caption{Photo of the sealed gaseous detector with CsI photocathode developed in collaboration with Oxford Instruments Inc.}
\label{oxford}       
\end{figure*}

\begin{figure*}
\centering
\resizebox{0.55\textwidth}{!}{
\includegraphics{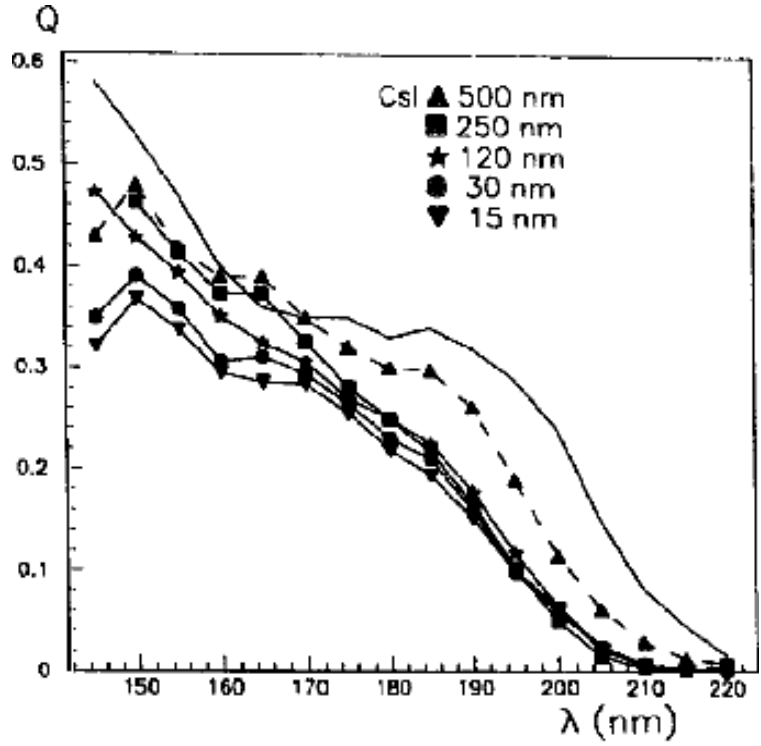}}
\caption{Quantum efficiency of a CsI photocathodes as a function of wavelength measured for various thickness \cite{Seguinot:1990uj}}
\label{QECsI}       
\end{figure*}

The principle of operation of a sealed single wire detector of a flame is illustrated in Fig. \ref{WorkingPrinciple} and its photo is shown in Fig. \ref{oxford}. UV light from the flame penetrates inside the detector through a transparent window and creates photoelectrons via photoelectric effect from the CsI photocathode. The typical quantum efficiency of reflective CsI photocathode manufactured at CERN is shown in Fig. \ref{QECsI}. The primary photoelectrons under the applied electric field drift towards the anode wire and initiate Townsend avalanches, which generate high amplitudes signals on the detector electrodes.

\begin{figure*}
\centering
\resizebox{0.65\textwidth}{!}{
\includegraphics{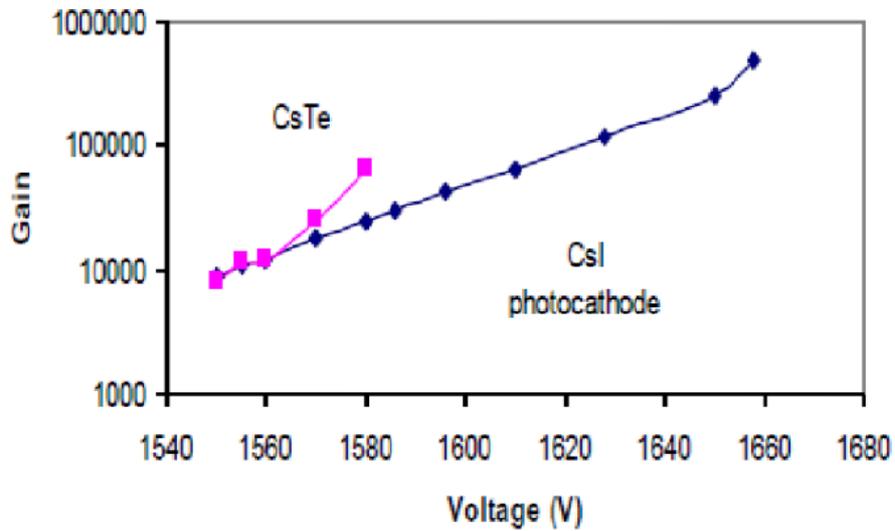}}
\caption{Gain as a function of voltage measured with two single-wire counters: one combined with a CsI and the other with a CsTe photocathode.}
\label{GainvsVoltage}       
\end{figure*}

\begin{figure*}
\centering
\resizebox{0.65\textwidth}{!}{
\includegraphics{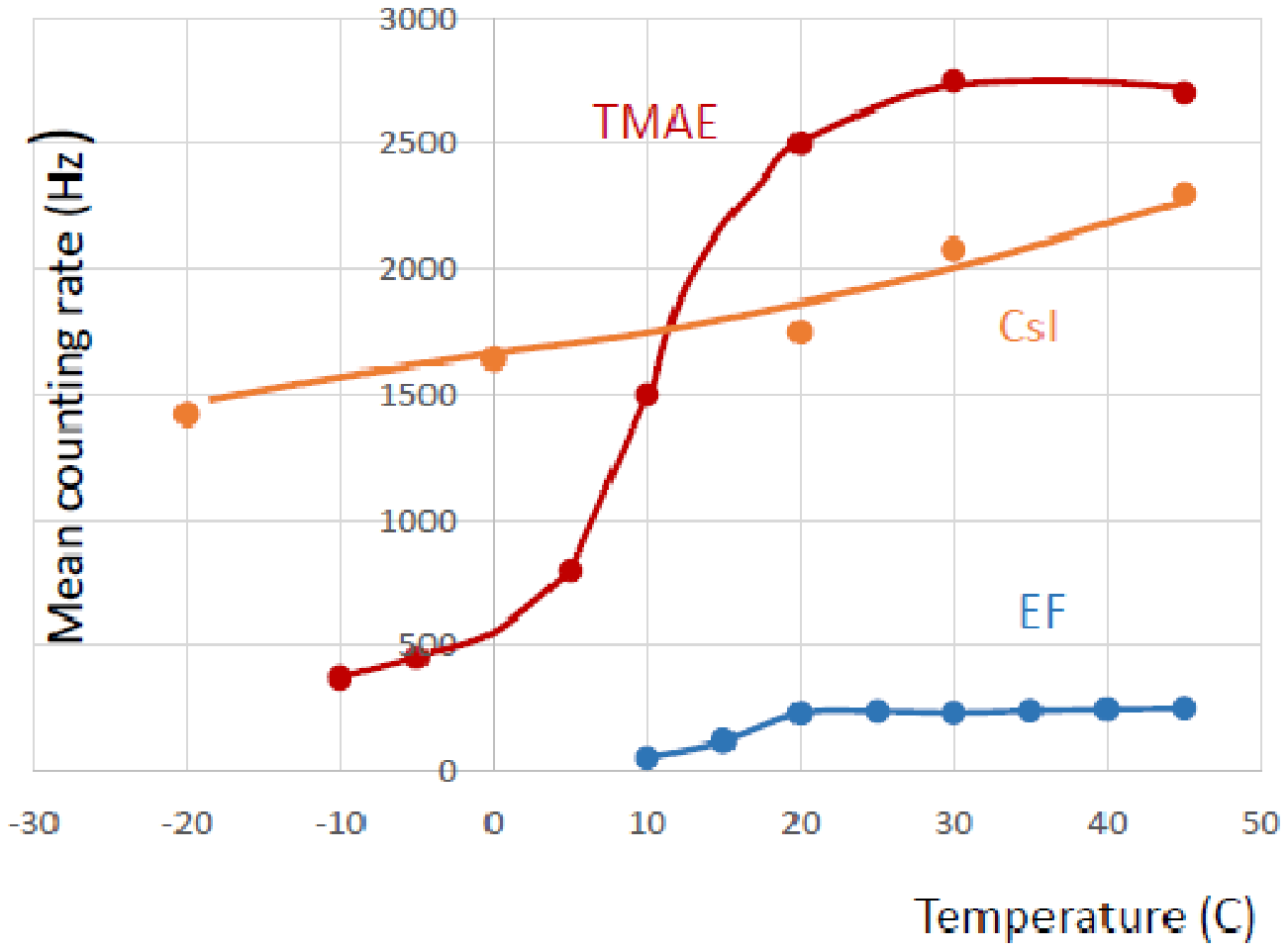}}
\caption{Single wire counters sensitivities as a function of temperature. Symbols CsI, TMAE an EF indicate: a detector with a CsI photocathode (see section \ref{sec:2.1}), a detector filled with TMAE vapour (see section \ref{sec:2.2.1}) and a detector filled with EF (see section \ref{sec:2.2.2}) respectively.}
\label{SensivsTemp}       
\end{figure*}

For the manufacturing of our first commercial prototype, a single wire counters production line at the Oxford Instrument (Finland), were used. Their standard detector, but without a window, was sent to CERN, where CsI was evaporated on the inner part of the cylindrical cathode on the surface located opposite to the hole for the window. Then it was put into the ordinary zip plastic bag filled with ambient air and sent back to the Oxford Instrument. Weeks later, after arrival to the company (all this time the detector was without the window and exposed to air) a quartz window was attached to the cathode cylinder and sealed to it.
The detector was then heated to 65$^\circ$C and stored under vacuum condition ($p$ $<$ 10$^{-6}$ Torr) for several days. After such a step, it was cooled to room temperature and filled with one of the gas mixtures: Ar+10\%CO$_2$ or Ar+10\%CH$_4$ at a total pressure of 1 atm. Its gain as a function of voltage is presented in Fig. \ref{GainvsVoltage}. As can be seen, the detector can operate at gains up to 2$\times$10$^5$. At higher gain the curve starts to deviate from the exponential behavior, indicating the appearance of photon feedback pulses. In Table \ref{table1} the comparative results of sensitivity measurements for three detectors are presented: Hamamatsu R2868, our industrial prototype and our laboratory prototype. 
As can be concluded from the presented data, our laboratory prototype is about 1000 times more sensitive and the commercial prototype is around 100 times more sensitive than Hamamatsu R2868. The lower sensitivity of the commercial prototype is connected to the fact that the CsI photocathode was exposed to ambient air for one week and it is known that a so long term contact with air cases its quantum efficiency degradation. Both detectors can be operated either in proportional or near Geiger mode. In proportional mode, in contrast to Hamamatsu R2868, our detectors can easily distinguish between a single photon and a short in time flux of photos (which produces a larger signal amplitude) and this allows reliably detect sparks. Note that time resolution in proportional mode was about 1 $\mu$s, which is a thousand times better than in the case of Hamamatsu R2868, allowing to record fast events and resolve them in time, such as sparks.
Although in a sealed gaseous detector the dependence $A$($V$) is not sensitive to the temperature, the quantum efficiency of the CsI slightly increases with the temperature (see Fig. \ref{SensivsTemp}). 
This effect was observed a long time ago \cite{osti_5232899}. However, it is not so important for the flame detection application, since as it is quite clear from Fig. \ref{SensivsTemp}, the detector has a high quantum efficiency in the entire temperature interval important from practice: between -20$^\circ$C and till +45$^\circ$C. 

A prototype of a single-wire detector with a CsTe photocathode, developed in a framework of collaboration with Reagent Inc., was tested as well. In principle, this photocathode offers a much better overlap between its quantum efficiency and the emission spectra of flames and thus looks as very perspective. However, in the gas mixture Ar+10\%CO$_2$, used in these preliminary studies, the gain curve starts to deviate from an exponential behavior at $A$ $>$ 2000. This is due to the appearance of photon feedback pulses disturbing the correct operation of the counter. The gain of about 2000 is marginal for the detection of primary single electrons with a higher efficiency even with low noise electronics. Therefore, in further work, it will be necessary to carefully optimize the gas mixture.

\begin{figure*}
\centering
\resizebox{0.7\textwidth}{!}{
\includegraphics{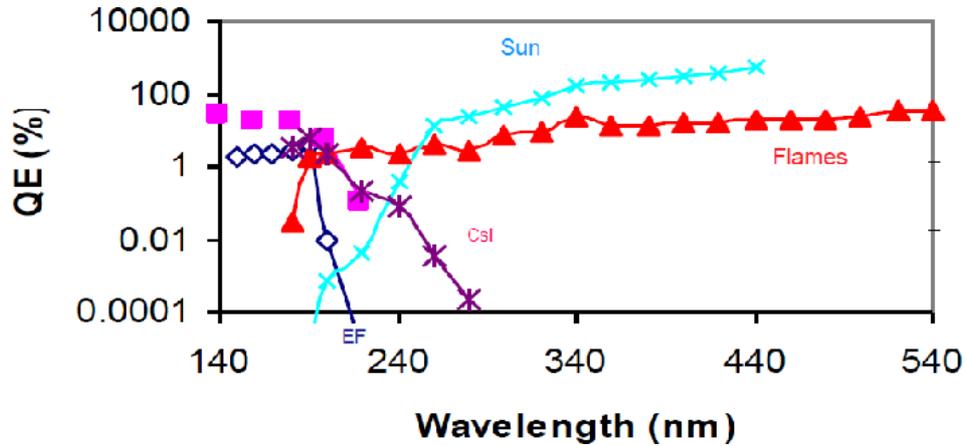}}
\caption{Results of the QE measurements for some detectors: stars$-$the QE of the single wire counter with the CsI photocathode and with the quartz window, filled squares$-$the QE of the same type of the detector, but with the MgF$_2$ window, open rhombus$-$the QE of the detector filled with the EF vapors. In the same figure are presented the typical spectra of flames in air (open triangles) and the spectra of the sun (open circles), both in arbitrary units \cite{hamamatsu, microtech}.}
\label{QEvsWL}       
\end{figure*}

Our test shows that in most indoor applications detectors with CsI photocathodes can record small flames in fully illuminated buildings without any filters. In contrast, a single wire with the CsTe photocathode, due to its much wider spectral range of sensitivity, suffering from background pulses caused by radiation with wavelengths $\lambda$~$>$~250 nm and thus requires filters to suppress the contribution from this radiation.
In outdoor applications, both detectors, with CsI and CsTe photocathodes, need filters to suppress undesirable pulses produced by the long wavelengths radiation (above ozone cut of 280 nm ) from the Sun. This is because the quantum efficiency of the CsI for $\lambda$~$>$~280 nm is not zero whereas the sun's emission in this spectral interval is extremely strong (see Fig. \ref{QEvsWL}). As a result, the convolution of the sunlight spectrum S($\lambda$) (in several photons penetrating the detector per d$\lambda$) with the CsI quantum efficiency Q($\lambda$) CsI in the spectral interval of 280$-$500 nm is not zero either and this gives a counting rate:

\begin{equation}
N_{\rm{CsI}} = \int_{280}^{500} S(\lambda)Q(\lambda)_{\rm{CsI}}\, d\lambda > 0
\label{eq1}
\end{equation}

The conclusion from these studies is that both CsI and CsTe photocathodes have very high efficiency and are attractive for indoor applications. They can be used also outdoor, but require filters to suppress undesirable pulses cases by the sunlight with wavelength the $\lambda$ $>$ 280 nm

\begin{figure*}
\centering
\resizebox{0.55\textwidth}{!}{
\includegraphics{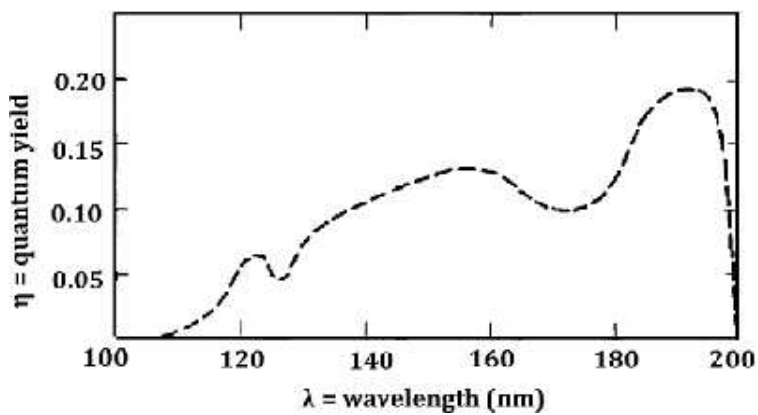}}
\caption{EF quantum efficiency as a function of photon wavelength \cite{Charpak:1988tu}.}
\label{QE_EF}       
\end{figure*}

\begin{figure*}
\centering
\resizebox{0.85\textwidth}{!}{
\includegraphics{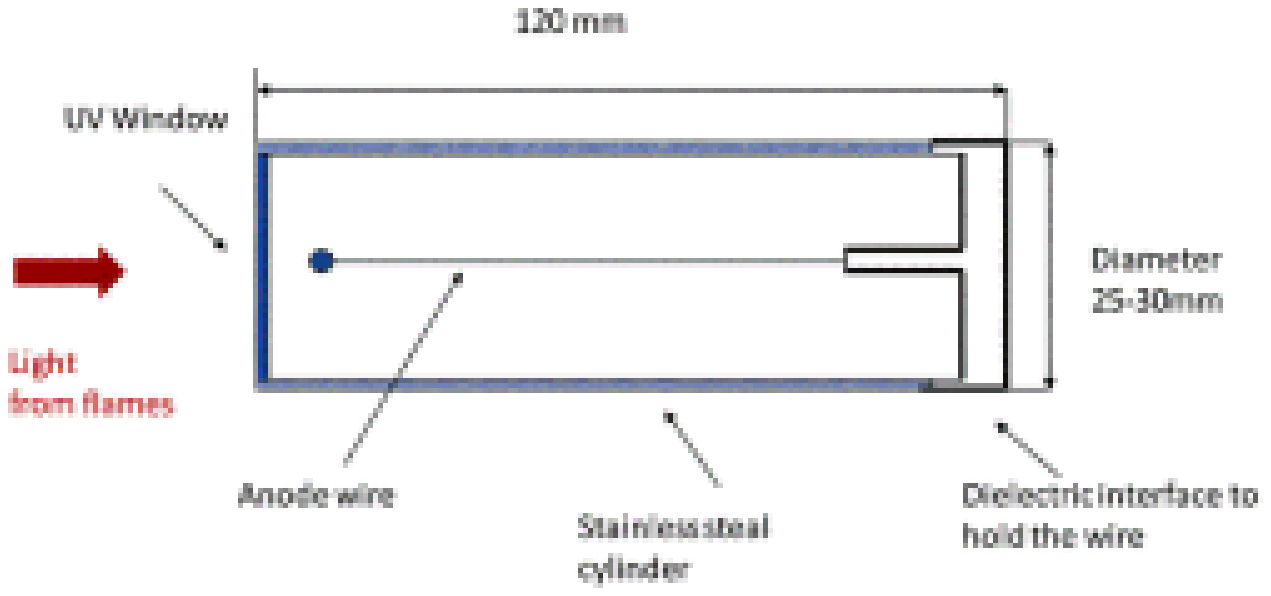}
\includegraphics{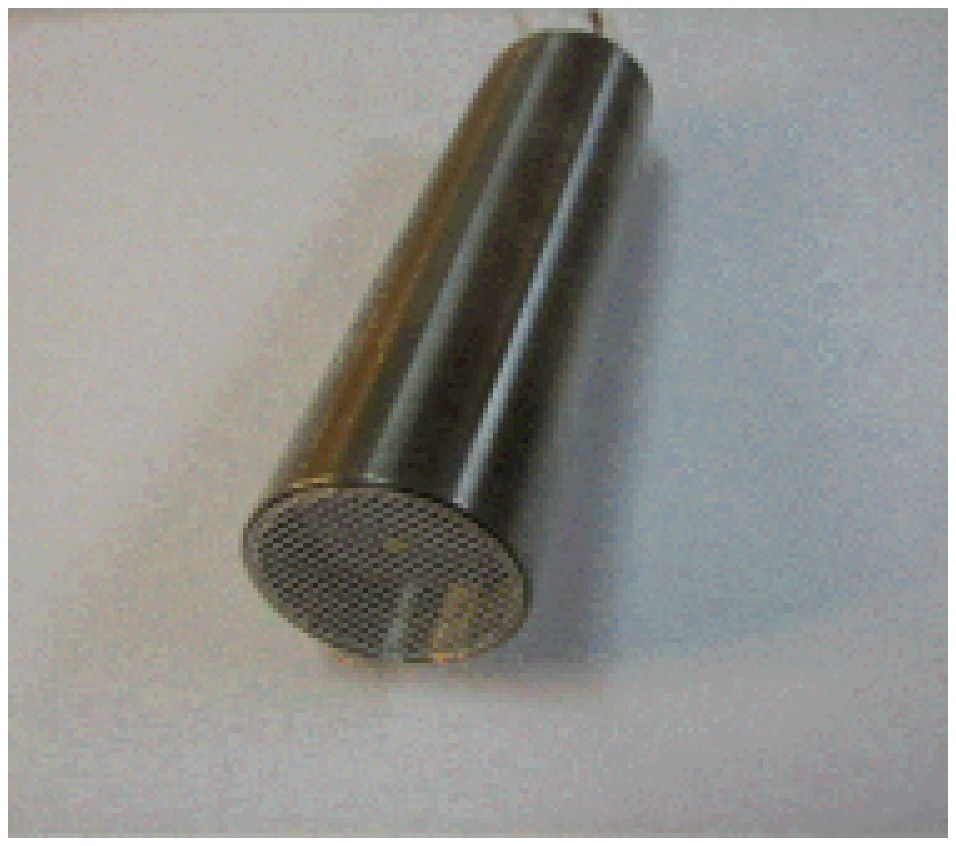}}
\caption{(Left) Schematics of a sealed single-wire flame detector filled with EF vapors. (Right) Its photograph.}
\label{SchemSealed}       
\end{figure*}

\begin{figure*}
\centering
\resizebox{0.7\textwidth}{!}{
\includegraphics{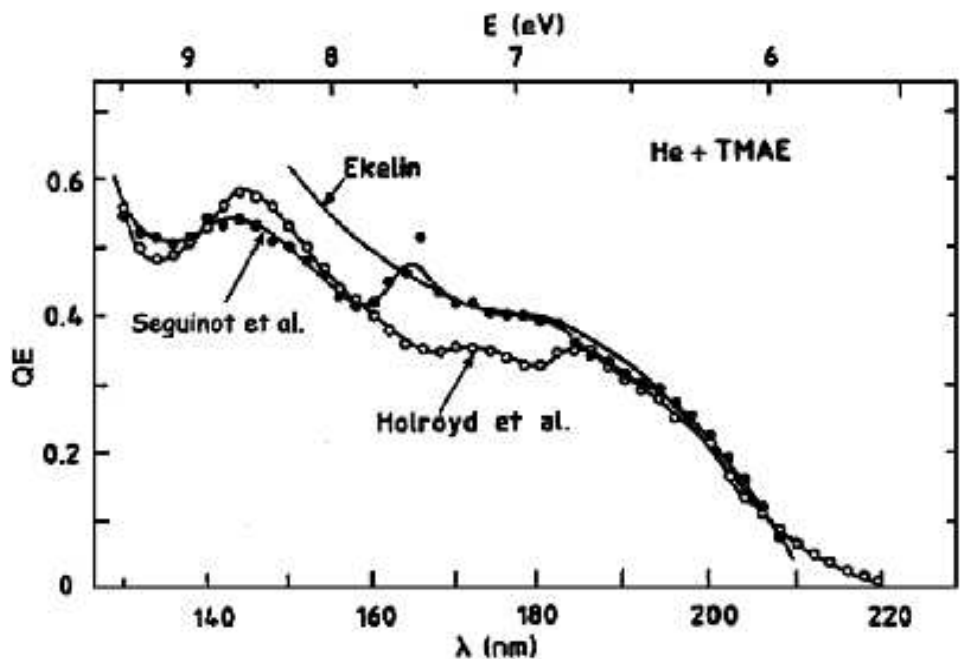}}
\caption{The quantum efficiency of TMAE vs wavelengths at full light absorption from \cite{Seguinot:1989ah}.}
\label{QE_TMAE}       
\end{figure*}

\subsection{Solar blind detectors filled with photosensitive vapours}
\label{sec:2.2}

The necessity to use the filters obviously affect the cost of the detectors.
The radical solution is to use photosensitive gases. In this case photoelectrons are created only when the photon energy $E$$_{\rm{\gamma}}$ $>$ $E_{\rm{i}}$, where $E_{\rm{i}}$ is the ionization potential of the gas. Consequently:

\begin{equation}
N_{\rm{psg}} = \int_{280}^{500} S(\lambda)Q(\lambda)_{\rm{psg}}\, d\lambda \approx 0
\label{eq2}
\end{equation}

Of courses, the gases to be used in the flame detector should have an ionization potential below 6.7 eV, corresponding to the ionization threshold of 185 nm. Several practical gases with such low ionization potentials were already identified earlier and were used (in flush mode only) in several applications. For example, single wire and multi-wire detectors flushed with ethylferrocene (EF) vapors ($E_{\rm{i}}$ = 6.2 eV) were used in plasma diagnostics \cite{https://doi.org/10.13140/rg.2.2.15133.20966}.
Tetrakis Dimethylamine Ethylene (TMAE) vapors, which has even smaller ionization potential $E_{\rm{i}}$ = 5.6 eV), were employed in some Ring Imaging Cherenkov detectors, e.g. DELPHI experiment at CERN \cite{Albrecht:1999ri}.
Therefore, the next focus of our work was on developing and test of sealed detectors filled with these gases.

\subsubsection{EF}
\label{sec:2.2.1}

The quantum efficiency of EF vapors under the condition of full absorption of the UV light is shown in Fig. \ref{QE_EF}. As can be seen, in the wavelength interval 140-195 nm its value is laying between 10 and 20\%, which is, in fact, as high as vacuum photomultipliers have. EF is a very convenient gas in practical use because it is not chemically aggressive and can be exposed to air. This dramatically simplifies its handling and consequently reduces the detector cost. However, it has low gas pressure and thus requires an extended absorption region in the detector to achieve the necessary efficiency at room temperature. For this reason, in our flame detector filled with EF vapors the entrance window is located at the end of the cylinder (see Fig. \ref{SchemSealed}), so that the light beam enters along the detector axis.

This detector was manufactured in the following way. First, the window with an anode wire attached to it was sealed to the end of the detector. The detector was then heated to 150$^{\circ}$C and pumped to a vacuum of 10$^{-6}$ Torr for several days. After the detector was cooled to room temperature and the EF vapors warmed to about 40-60$^{\circ}$C were introduced (to ensure their condensation). Finally, the detector was filled by Ar+15\%CO$_2$ at a to total pressure of 1 atm. In Fig. \ref{SensivsTemp}, the efficiency of the EF vs the temperature is shown. As can be seen, at temperature $t$ $>$20$^{\circ}$C all EF condensed inside the detector during the manufacturing procedure is evaporated and the efficiency reaches the plateau. Of course, the exact behavior of this curve depends on the amount of the condensed EF. Comparison of sensitivities of CsI and EF photocathodes are presented in Fig. \ref{QEvsWL}. As can be seen, compared to CsI, EF has a sharp cut of the sensitivity at wavelengths more than 200 nm, making it a solar blind. Measurements show that when direct sunlight penetrates into the detector the counting rate was $\approx$ 0.3 Hz and it was mainly caused by the cosmic radiation. With such a low background it was easy to detect small flames even in the presence of direct sunlight. Comparison to commercial sensors of UV radiation for flames able to operate in sunlight conditions (for example, Hamamatsu UVtron R2868) shows that the sensitivity of our detector is at least 100 higher whereas the estimated price can be a few times lower than for example PM based detectors (see Table \ref{table2}). The conclusion from these tests is that detectors filled with EF are attractive for outdoor applications only in countries with warm climates, such as Greece, Italy etc.

\subsubsection{TMAE}
\label{sec:2.2.2}

\begin{figure*}
\centering
\resizebox{0.7\textwidth}{!}{
\includegraphics{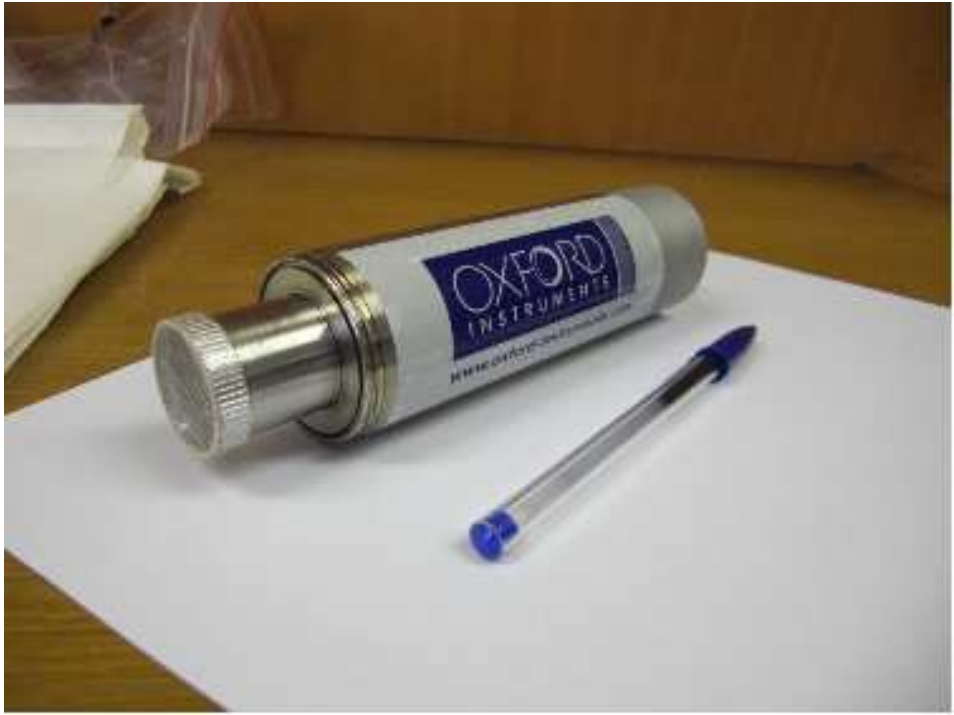}}
\caption{Sealed single wire counter filled with TMAE vapors.}
\label{sealedTMAE}       
\end{figure*}

The main advantage of TMAE is that they offer higher quantum efficiency, which is very close to the CsI one (compare figures \ref{QECsI} and \ref{QE_TMAE}) as well as higher vapor pressure than EF.
For these reasons TMAE vapors were earlier used in several RICH detectors, for example in DELPHI experiment.
However, TMAE is chemically aggressive: it interacts with many materials, including air, so in contrast to EF it is much more difficult to handle. For this reason, it was quite challenging to make a sealed detector filled with TMAE vapors. This required a careful choice of materials (such as stainless steel, glass, quartz, and ceramics) and a more complicated system for gas filling allowing heating to a high temperature to remove, if necessary, TMAE contamination.
The sealed single-wire detector was manufactured similarly to one filled with EF, however, its design was slightly changed. The problem is that TMAE absorbed on the cathode surface increases its quantum efficiency at least on one order of magnitude in a wide range of the spectrum, including a visible region \cite{Peskov:1987kr}. For this reason, the detector has a collimator preventing the entering light to directly heat the cathode (see Fig. \ref{sealedTMAE}).
The temperature dependence of its quantum efficiency is shown in Fig. \ref{SensivsTemp}. As can be seen, it increases with temperature and at $t$ $>$ 20$^{\circ}$C reaches a kind of saturation, indicating that all excess TMAE condensed inside the detector was evaporated. The efficiency of the TMAE detector in the ``plateau" region is approximately 10 times higher than the one filled with EF.
The background counting rate caused by the direct sunlight was about 17 Hz. This was 5 times higher than in the case of our laboratory prototype. The possible explanation is that TMAE vapors interacted with a glue used to attached anode wire to the entrance window and the contaminate by this substance gas, adsorbed on the detector cathode, causes enhancing of its quantum efficiency.
In Table \ref{table2} a comparison between the sensitivities of sealed single-wire counters filled with EF and TMAE vapors is given. At the same table, as a reference, the sensitivity of the Hamamatsu sensor is presented as well. As can be seen, the detector with TMAE almost 1000 more sensitive than R2868 and 10 times more sensitive that the detector filled with EF vapors.
Another important feature of detector using TMAE,  as well as that filled with EF, is that in contrast to the Hamamatsu sensor, it operates in proportional mode and has also a superior time resolution about $\approx$$\mu$s. In proportional mode, from the pulse amplitude and the time dependence, one can easily distinguish between sparks in the monitoring area, producing many photons in a short period and single photons.
In case, when the mean spark rate is below the counting rate caused by the cosmic rays or the background caused by the direct sunlight, very good results where obtained with two TMAE detectors operating in parallel. In signal coincidence mode one can reject pulses produced by cosmic rays and achieve sensitivity to sparks in the monitoring area as high as in Table \ref{table2} even in direct sunlight conditions. Thus such detector could be used either in an indoor or outdoor application, however the later case in countries with a warm climate. This, of course, imposed some restrictions on applications.

\begin{table}[]
\centering
\caption{Comparison of sensitivities of Hamamtsu R2868 with our sealed flame detectors filled with EF and TMAE vapors. All measurements were performed with a candle in a fully illuminated building.}
\begin{tabular}{|l|l|l|l|l|l|}
\hline
\multicolumn{2}{|c|}{\textbf{Hamamatsu R2868}}                                                & \multicolumn{2}{c|}{\textbf{Our TMAE detector at 23$^{\circ}$C}}                                                                                   & \multicolumn{2}{c|}{\textbf{Our EF detector at 25$^{\circ}$C}}                                                                                       \\ \hline
Distance (m) & \begin{tabular}[c]{@{}l@{}}Mean number of\\ counts per 10 sec: $N_{\rm H}$\end{tabular} & \begin{tabular}[c]{@{}l@{}}Mean number of\\ counts per 10 sec: $N_{\rm TMAE}$\end{tabular} & \begin{tabular}[c]{@{}l@{}}Ratio \\ $N_{\rm TMAE}$/$N_{\rm H}$\end{tabular} & \begin{tabular}[c]{@{}l@{}}Mean number of\\ counts per 10 sec: $N_{\rm EF}$\end{tabular} & \begin{tabular}[c]{@{}l@{}}Ratio \\ $N_{\rm EF}$/$N_{\rm H}$\end{tabular} \\ \hline
1.1          &                      583                                                    &                             690747                                    &              1.18$\times$10$^3$           &                      75613                                                  &            1.20$\times$10$^2$                                            \\ \hline
3.0          &                        76                                                     &            91013                                                           &        1.19$\times$10$^3$              &                 11052                                                      &         1.40$\times$10$^2$                                             \\ \hline
10.0         &                           6                                                   &             7820                                                             &     1.30$\times$10$^3$                &               643                                                             &         1.20$\times$10$^2$                                             \\ \hline
30.0         &                     0.1                                                           &         873                                                                          &       8.0$\times$10$^3$     &                68                                                              &        6.00$\times$10$^2$                                              \\ \hline
85.0         &                           0.0                                                     &             51                                                                      &                                                      &             4                                                                    &                                                      \\ \hline
\end{tabular}
\label{table2}
\end{table}

\section{Sealed large sensitive area flat panel detectors}
\label{sec:2}

An alternative approach is to use a flat panel designs for spark and flame detection. The main advantage is that one can increase the sensitivity of the detector proportional to its sensitive area. Note that Hamamatsu R2868, due to the feature of glow discharge, has a limit on its sensitive surface around 1 cm$^2$.  In this work we tested two designs, allowing to easily increase the sensitive area: Multi-Wire Proportional Chambers (MWPCs) and GEM-based detectors.

\subsection{Flat panel multi-wire detectors}

\begin{figure*}
\centering
\resizebox{0.5\textwidth}{!}{
\includegraphics{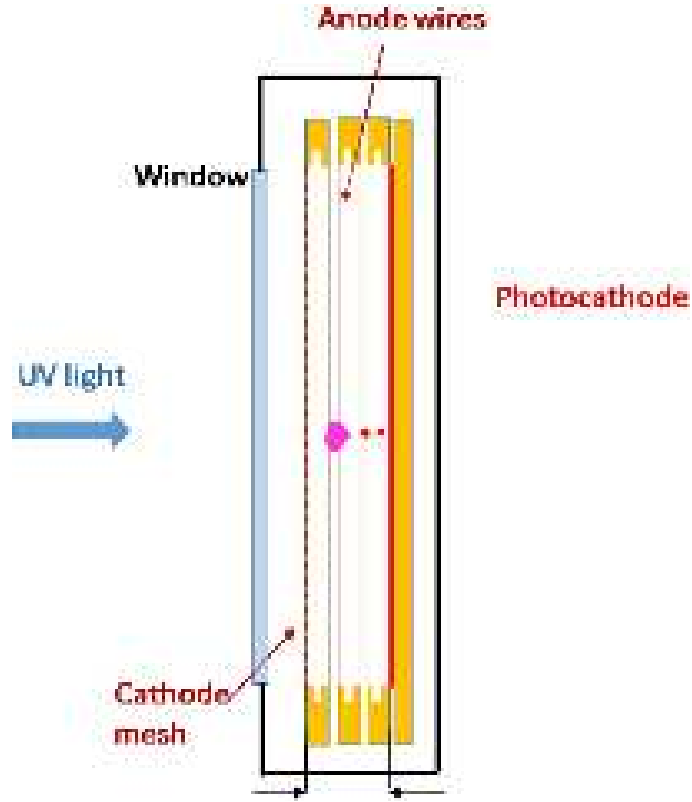}}
\caption{Sealed single wire counter filled with TMAE vapors.}
\label{fig_a}       
\end{figure*}

\begin{figure*}
\centering
\resizebox{0.5\textwidth}{!}{
\includegraphics{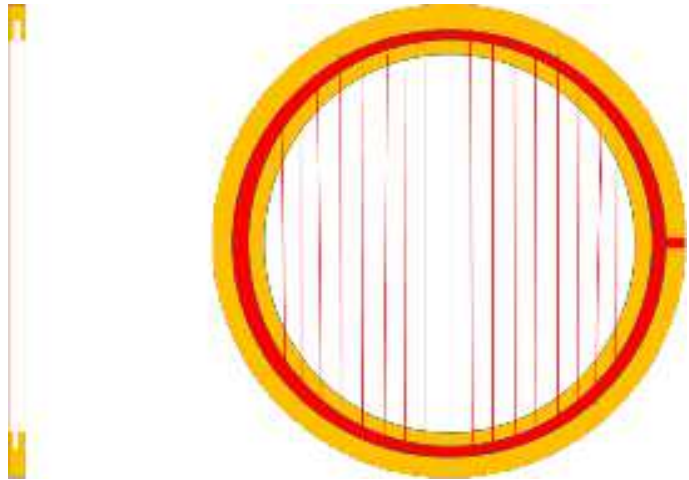}}
\caption{Anode plane.}
\label{fig_b}       
\end{figure*}

\begin{figure*}
\centering
\resizebox{0.5\textwidth}{!}{
\includegraphics{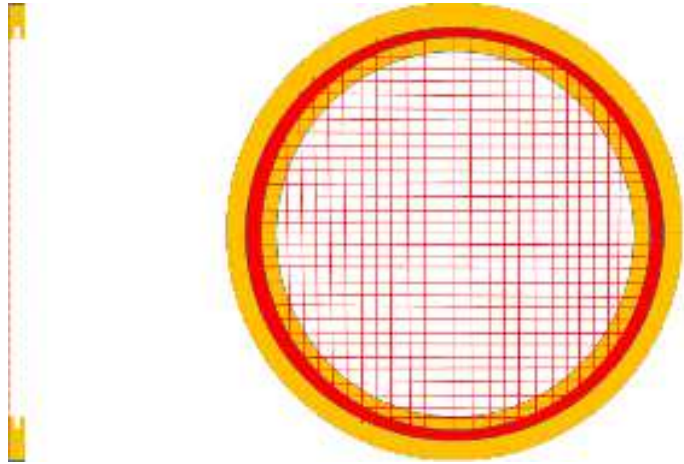}}
\caption{Ring with a stretched mesh.}
\label{fig_c}       
\end{figure*}

\begin{figure*}
\centering
\resizebox{0.5\textwidth}{!}{
\includegraphics{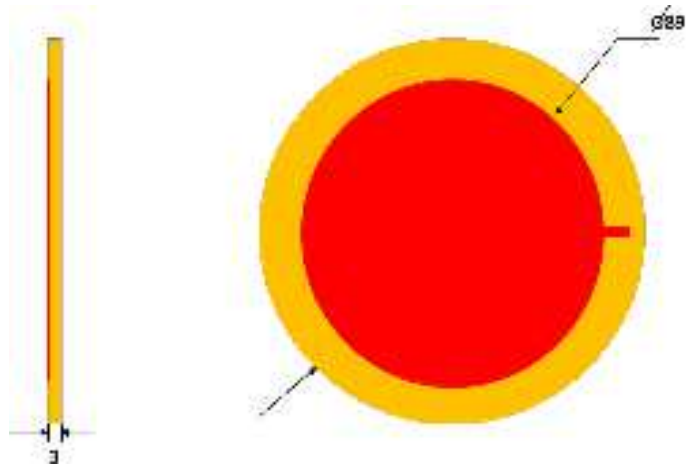}}
\caption{Disc with photocathode.}
\label{fig_d}       
\end{figure*}

The ALICE experiment at CERN developed and are using MWPCs combined with CsI photocathodes for detection of Cherenkov photons \cite{Hoedlmoser:2006ra}. These detectors operated at a gas gain of about 5$\times$10$^4$, have high quantum  efficiency in UV region spectra and are able to detect single photoelectrons with an efficiency close to 95\%. 
Coming from this experience we developed a miniature version of this device for detection of sparks and flames. The design of this MWPC is shown schematically in Fig. \ref{fig_a}. It consists of two G-10 rings (Fig.~\ref{fig_b} and Fig.~\ref{fig_c})  and a G-10 disc (Fig. \ref{fig_d}). The latter had a metallized surface and served as photocathode substrate. Two other were supporting frames for stretched anode wires (Fig.~\ref{fig_b}) and the upper cathode mesh  (Fig.~\ref{fig_c}).

\begin{figure*}
\centering
\resizebox{0.7\textwidth}{!}{
\includegraphics{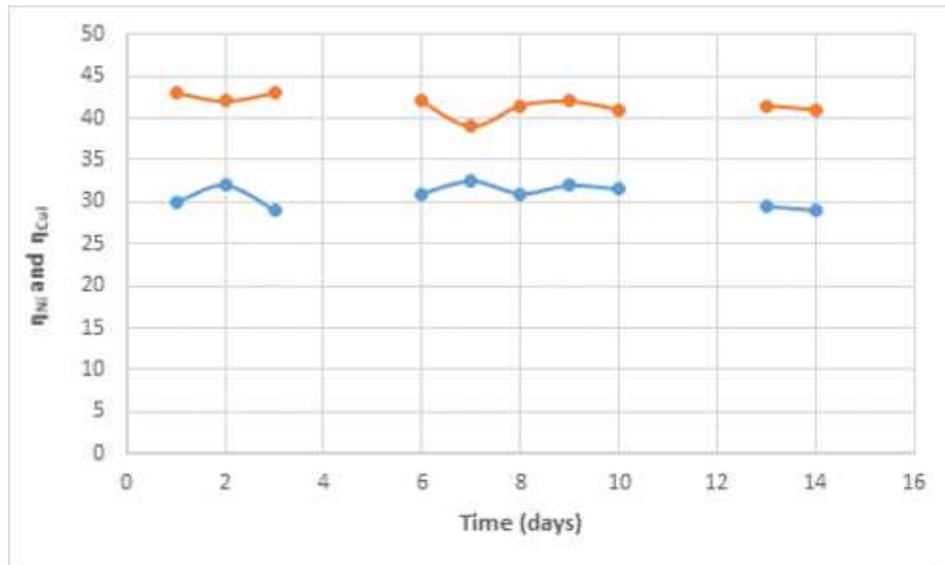}}
\caption{Relative efficiencies of sealed MWPCs combined with Ni and CuI photocathodes vs time.}
\label{fig_e}       
\end{figure*}

The MWPC detectors were mounted inside a stainless steal test chamber, having a quartz window and feedthroughs for supplying detectors with  HV and taking signals from them (Fig. \ref{fig_a} and Fig. \ref{fig_b}). 
For the system outgassing the chamber and the appropriate part of the gas system was heated to $\approx$ 150$^{\circ}$C  with electric tapes wrapped in aluminum foil (in order the heat uniformly). Under these conditions  the chamber was pumped for several days for vacuum better than 10$^{-6}$ Torr. After that Ar+20\%CO$_2$ gas mixture at pressure 1 atm was introduced.
Although preliminary measurements were done with a CsI photocathode, the main focus of this work was on tests of solid solar-blind photocathodes: Ni and CuI, because in latter case no filter are needed for outdoor applications. They have, of course, much less sensitivity than CsI photocathode, however one compensate this by the larger surface. Moreover, due to the reduced photon feedback effect (compared to the CsI option), one can operate MWPC at rather high gas gains 1-3$\times$10$^5$ without any breakdowns, which allowed to achieve almost  a 100\% detection efficiency for the photoelectrons detection.
The result of the brie tests of these detectors in sealed mode are shown in Fig. \ref{fig_e}, in which relative efficiencies $\eta_{\rm{Ni}}$ and $\eta_{\rm{CuI}}$  of MWPCs combined with Ni and CuI photocathodes respectively, are shown. The relative efficiencies are define as follows:

\begin{equation}
\eta_{\rm{Ni}} = N_{\rm{Ni}}/N_{\rm{Ham}} \hspace{0.8cm} \eta_{\rm{CuI}} = N_{\rm{CuI}}/N_{\rm{Ham}}
\label{eq3}
\end{equation}

where $N_{\rm{Ni}}$, $N_{\rm{CuI}}$  and $N_{\rm{Ham}}$ are counting rate measured with Ni and CuI MWPCs and Hamamatsu R2868 when the candle was placed at a distance of 2 m from them.

Manufacturing even small size MWPCs is a time consuming process requiring well training technicians. Therefore, one can expect that commercial flame detectors based on multi-wire structures will be more expensive than the single wire options.
Taking this into account we developed a conceptional design of a simplified version of a commercial prototype of a large area flame detector.
 It consists from an array of single wire counters with rectangular Ni photocathodes. The main simplification was that only four wires were used which were stretched and soldered on a standard printed circuit board (PCB). This makes it much easier to manufacture than ether single wire or small pitch MWPC described above. The active area of 8$\times$8 cm$^2$ makes it 64 times more sensitive than Hammatsu R2868.  Another advantage of this approach is its  larger angle acceptance which a favorable factor for some applications.

\begin{figure*}
\centering
\resizebox{0.7\textwidth}{!}{
\includegraphics{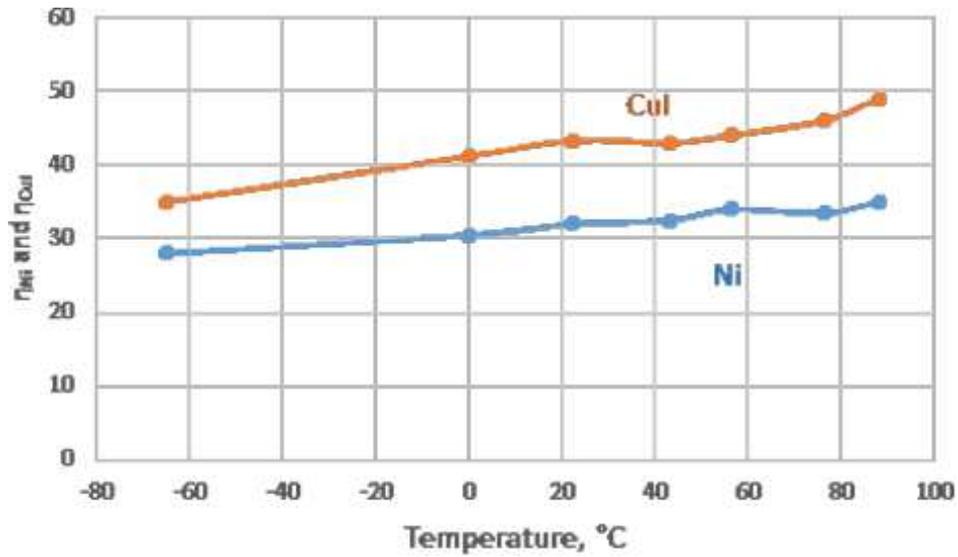}}
\caption{Relative efficiencies of sealed MWPCs photocathodes vs temperature.}
\label{fig_f}       
\end{figure*}

\begin{figure*}
\centering
\resizebox{0.7\textwidth}{!}{
\includegraphics{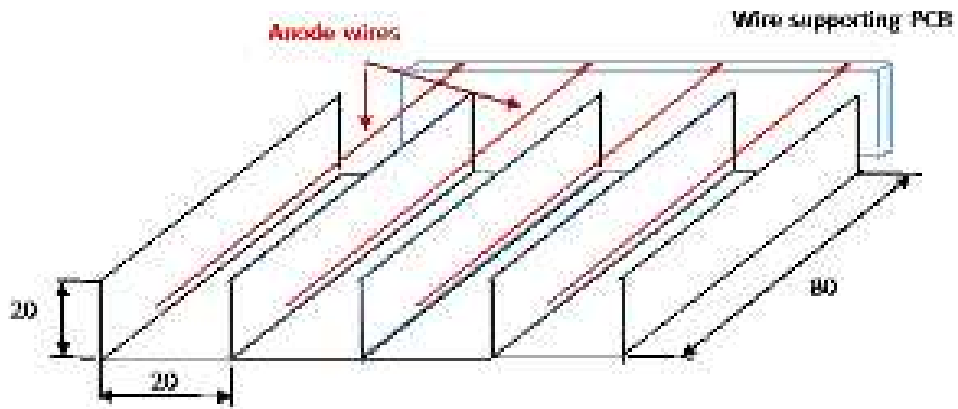}}
\caption{Schematic design of an array of rectangular single wire counters with Ni photocathodes.}
\label{fig_g}       
\end{figure*}

\subsection{Sealed GEM-based detectors}

An estimated cost of mass production of sealed single wire gaseous detector described above is around 100 Euro including the cost of a battery feed electronics if there is already producing a line. This estimation was performed independently by the experts from the Oxford Instr., Finland and DIEHL, Germany.
Unfortunately, the production of single-wire counters in the world practically does not exist anymore, in most applications they were replaced by solid-state detectors, MWPCs are mainly manufactured in laboratories.
In this situation an effective approach will be to use other type of multiplication structures, which are commercially available, for example Gas Electron Multipliers (GEMs) \cite{Sauli:1997qp}.
Potential advantages of the GEM-based flame detectors are: they offer compact flat-panel type design, they may have large sensitive areas and the possibility to operate in wide range of gases, including pure noble gases, and finally, they have imaging capability.
There were already attempts to develop sealed gaseous photomultipliers based on hole-type tape amplification structures. For example, sealed GEM detector combined a CsI photocathode demonstrated an excellent  stability in time \cite{Breskin:2002sz}.
Moreover, as was shown in our works \cite{Giunji:2002dj} even more sophisticated photocathodes remain stable in sealed gaseous photomultipliers, if the latter are made of properly chosen materials such as glass, ceramics, etc.).
 
In this work, sealed flame detectors, based on hole-type multipliers, so called thick GEM (TGEM) \cite{Periale:2001ya} or Resistive GEM (RGEMs) \cite{Bidault:2006kx} were used due to their robustness and low cost in mass production (one euro for 1000 holes). These electron multipliers were made of a printed circuit boards (a mixture of a fibers glass or sometime even paper,  with an epoxy resin) and thus, in contrast to Kapton or glass, have a lot cavities and contain strongly outgassing materials. Therefore, the first task was to investigate if such multipliers could operate stably in sealed mode.  In gas flush mode they were already used for detection of Cherenkov photons \cite{Martinengo:2011zz,Peskov:2011ei,Alexeev:2012tba}. In the following, results of systematic studies of operation of sealed flame detectors based on TGEM and RETGEM are reported. 

\subsection{Setup}

\begin{figure*}
\centering
\resizebox{0.7\textwidth}{!}{
\includegraphics{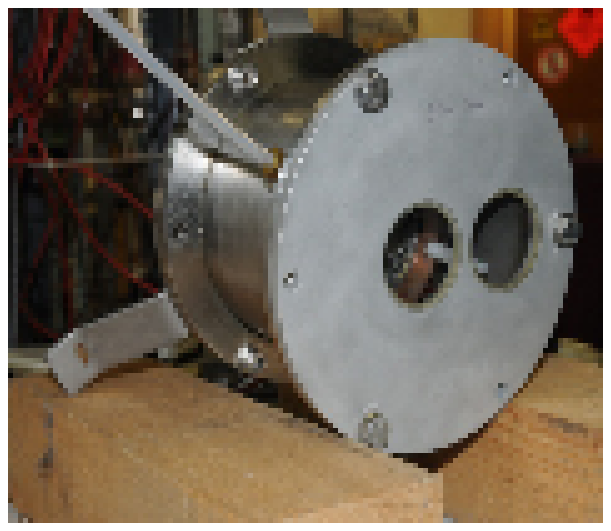}
\includegraphics{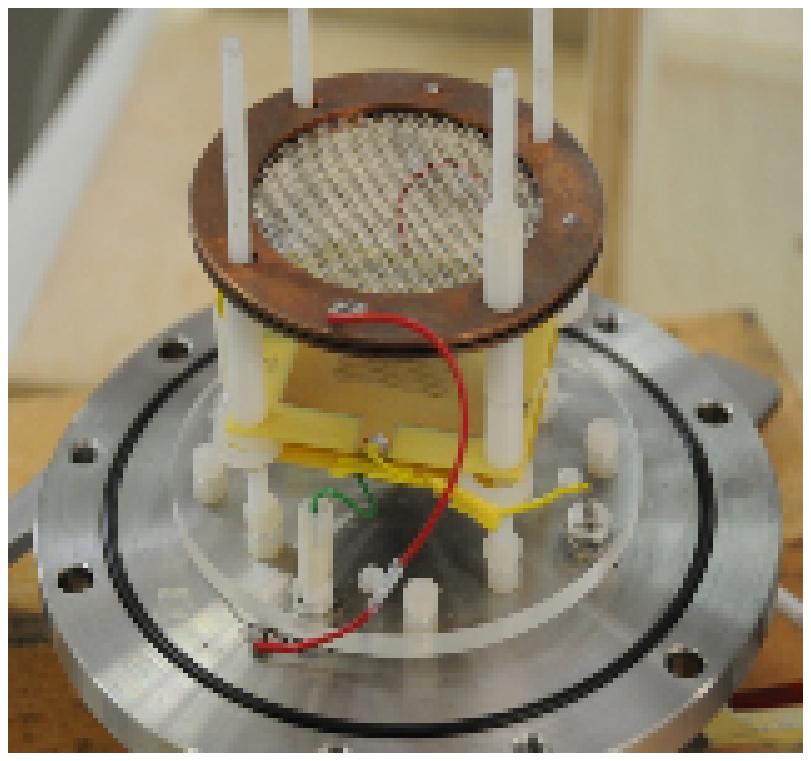}}
\caption{Photograph of the chamber for the stability tests of TGMs and TGEMs operating in sealed mode (left) and its inner part (right).}
\label{GEMsealed}       
\end{figure*}

\begin{figure*}
\centering
\resizebox{0.6\textwidth}{!}{
\includegraphics{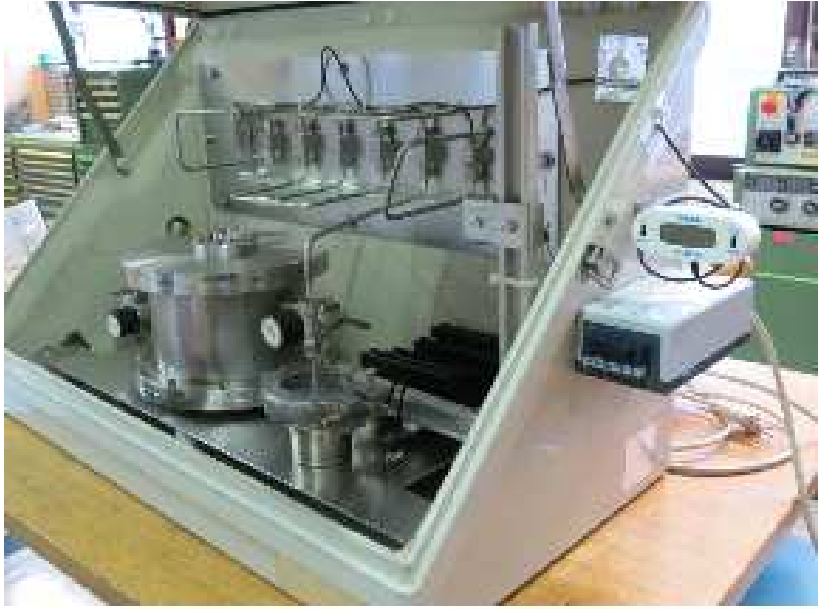}}
\caption{Photograph of a setup for outgassing and filling with the gas of sealed gaseous detectors.}
\label{sealedSetup}       
\end{figure*}

TGEM or RETGEM detectors were mounted inside a stainless steel test chamber, having a quartz window and feed-throughs for supplying detectors with  HV and taking signals from them (Fig. \ref{GEMsealed}). Most of the detectors supporting structures were made of ceramics
In the first experiment, a quite simple approach was used for the system outgassing: the chamber and the appropriate part of the gas system was heating to 150-80$^{\circ}$C  with electric tapes wrapped in aluminum foil (in order the heat uniformly). Under these conditions, the chamber was pumped for 7$-$10 days for vacuum better than 10$^{-6}$ Torr.
Later we prepared and start using a special setup shown in Fig. \ref{sealedSetup}. It was a metallic cabinet accommodating the detector and the gas system including bottles filled photosensitive liquids. The temperature inside was regulated with a heating element and maintained automatically at the set value. The temperature uniformity inside the cabined was ensured with two vents.

\subsection{Results of the tests}
\label{sec:2.2}

\begin{figure*}
\centering
\resizebox{0.8\textwidth}{!}{
\includegraphics{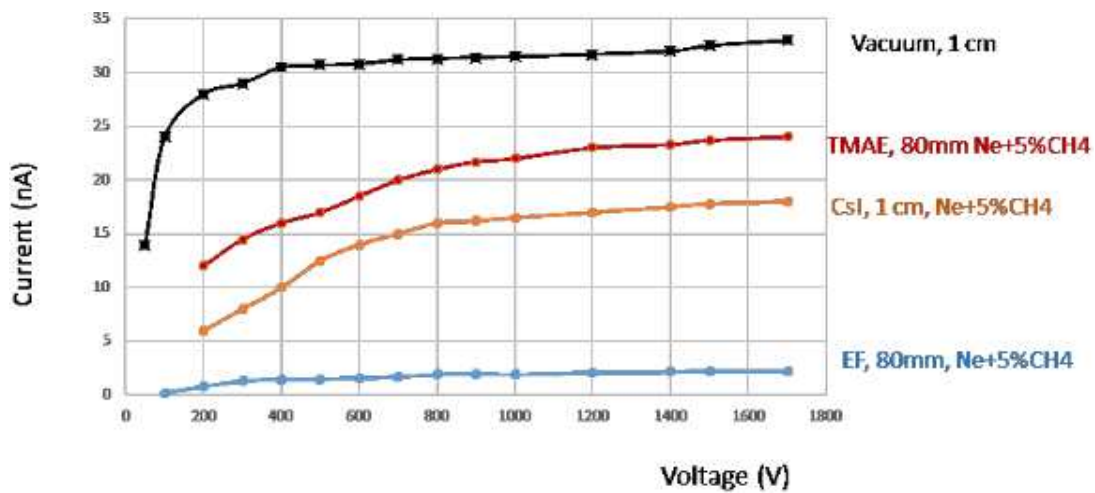}}
\caption{Results of measurements of relative sensitivities of sealed photosensitive gaseous detectors at 185 nm. In the case of CsI the length of the drift region was 10 mm. In the case of  photosensitive gases it was 80 mm. Measurements indicated as $Vacuum$ were performed in vacuum 10$^{-6}$ Torr with the CsI photocathode; the loss of the quantum efficiency in the gas was due to the electrons back scattering effect \cite{PhysRev.40.980,doi:10.1063/1.1721223}.}
\label{sensyGEM}       
\end{figure*}

\begin{figure*}
\centering
\resizebox{0.6\textwidth}{!}{
\includegraphics{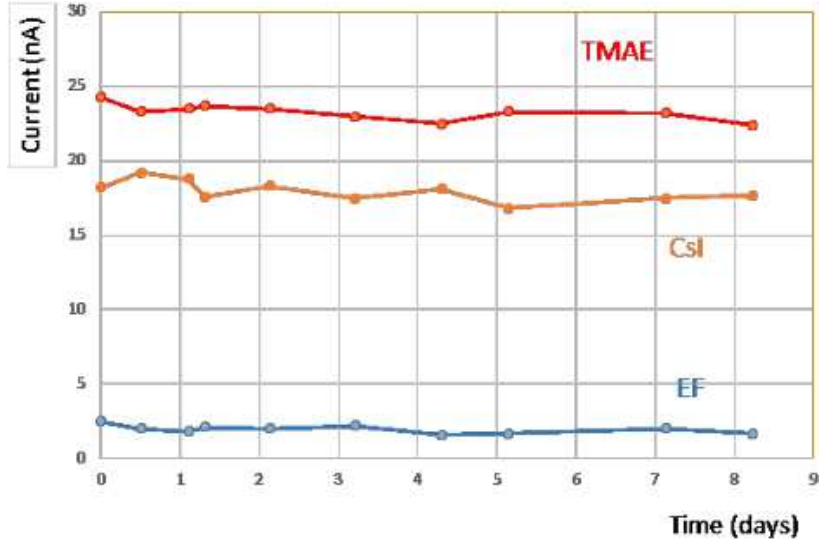}}
\caption{Photocurrent vs. time, measured with CsI, TMAE and EF photocathodes.}
\label{currentvstimeGEM}       
\end{figure*}

After the chamber outgassing, the desired gas was introduced. In most of the studies, Ne or mixtures of Ne with a small amount of CH$_4$ or CO$_2$ were used, which offered lower operational voltages than other noble gases for the same gas gain. First measurements were done in the ionization chamber mode (Fig. \ref{sensyGEM}). This allowed us to evaluate relative quantum efficiencies of the photocathodes and to monitor their stability in time (Fig. \ref{currentvstimeGEM}).

\begin{figure*}
\centering
\resizebox{0.9\textwidth}{!}{
\includegraphics{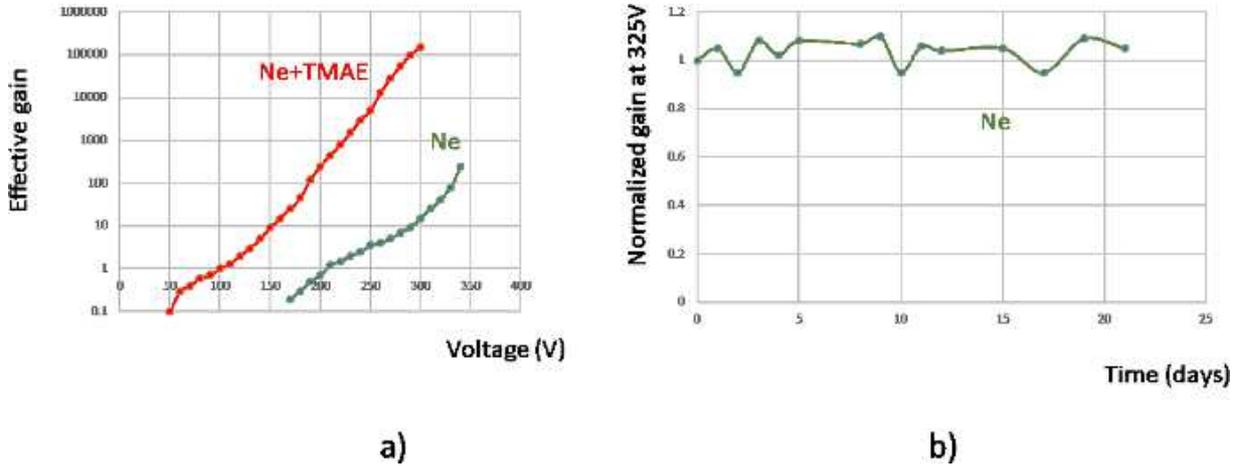}}
\caption{(a) Gas gain of a double TGEM  ($t$ = 0.4, $d$ = 0.5, $s$ = 0.9, $h$ = 0.1 mm) in pure Ne and in Ne+TMAE gas mixture and (b) gain stability in Ne vs. time at $A$ = 50.}
\label{gasGainDoubleTGEM}       
\end{figure*}

In the next series of experiments, we checked the gas gain stability for these sealed GEM-based detectors. First such measurements were performed in pure Ne (Fig. \ref{gasGainDoubleTGEM}) at a pressure of 1 atm because the gain value at the fixed voltage this gas is very sensitive to tiny impurities. Therefore, Ne was a very convenient gas to monitor possible outgassing of the gas chamber. 
However, long-time outgassing causes TGEMs over-drying so that some TGEMs charging up effect was observed.
In mixtures of Ne with quenched gases, good long-term gain stability was observed at low counting rate even after 1-2 days of heating and pumping of the gas chamber and the short term charging up effect was less pronounced, probably because not all moisture was removed from G-10 during this relatively short time of pumping.
This is not the case when mixtures with TMAE are used. In this case, TGEMs become very noisy and after few hours a leak current of few nA, and sometimes even more, appears.
We assumed it was due to the TMAE absorption in the G-10 (fiberglass) plate, from which TGEMs were manufactured.  To reduce this effect we introduced TMAE vapors at their pressure less than the saturated value at room temperature (which is about 0.4~Torr at 20$^{\circ}$C). This increases the time before TGEMs become noisy and reduces the leak current, but the efficiency corresponding dropped. 
Another efficient way was to keep the detector at some elevated temperature, for example, 40$^{\circ}$C, but of course, the entire system becomes more complicated and correspondingly more expensive. 
The problem with the leak current stimulated us to develop special TGEM designs which we call wall-less. Two prototypes were built and tested.

\subsection{Wall-less TGEM}

\begin{figure*}
\centering
\resizebox{0.6\textwidth}{!}{
\includegraphics{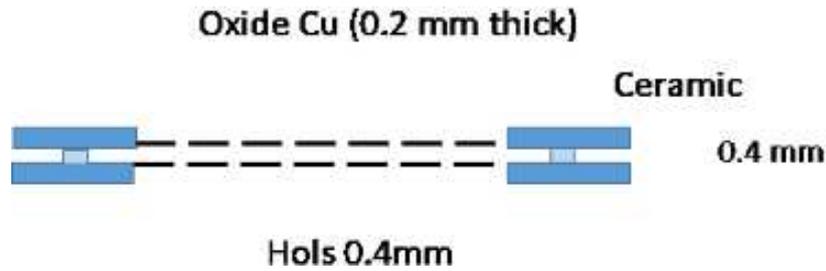}}
\caption{Schematic drawing of the $wall-less$ TGEM with suspended copper oxide electrodes having  aligned holes  and spacers located outside the active area.}
\label{wall_less}       
\end{figure*}

\begin{figure*}
\centering
\resizebox{0.7\textwidth}{!}{
\includegraphics{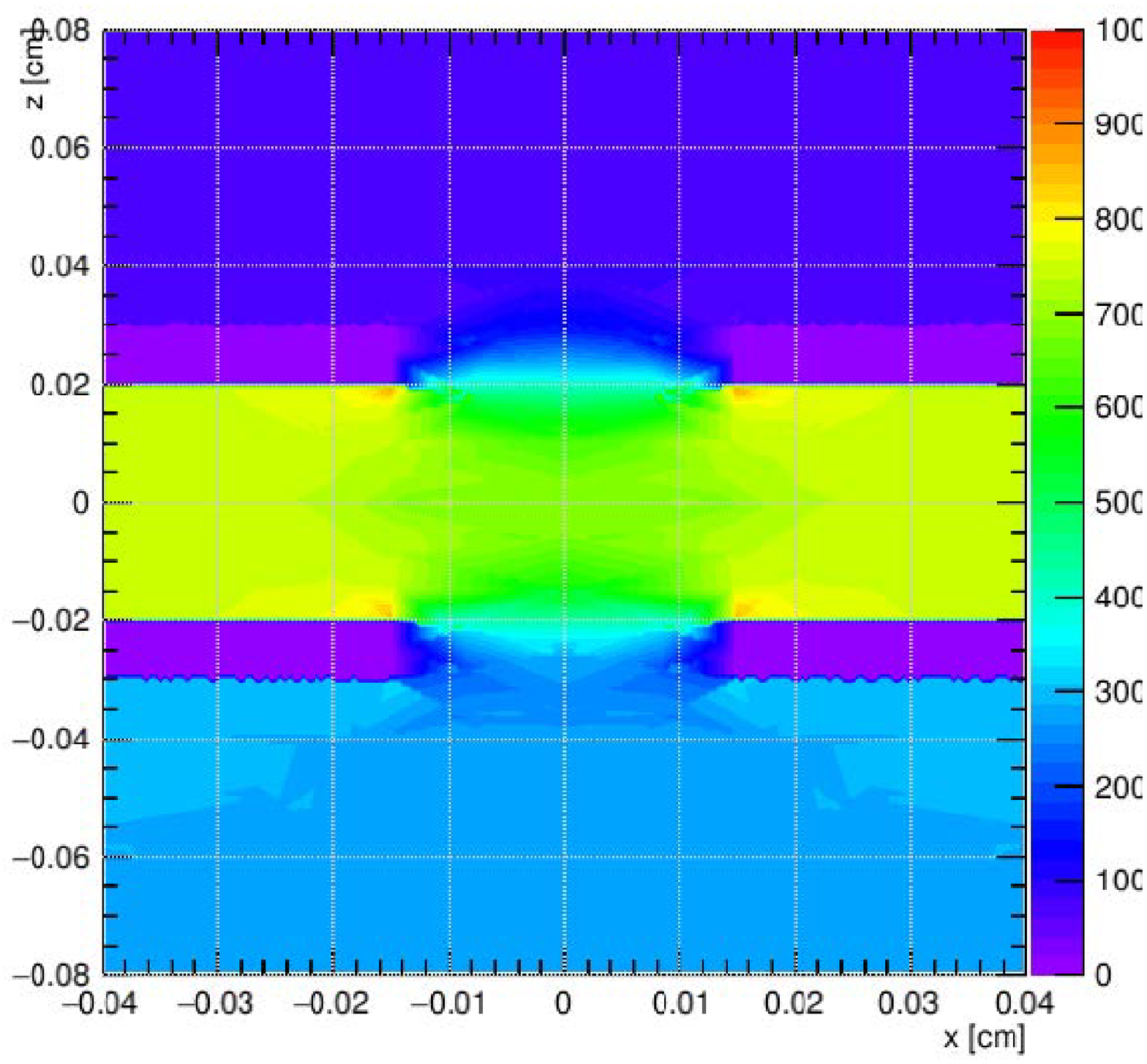}
\includegraphics{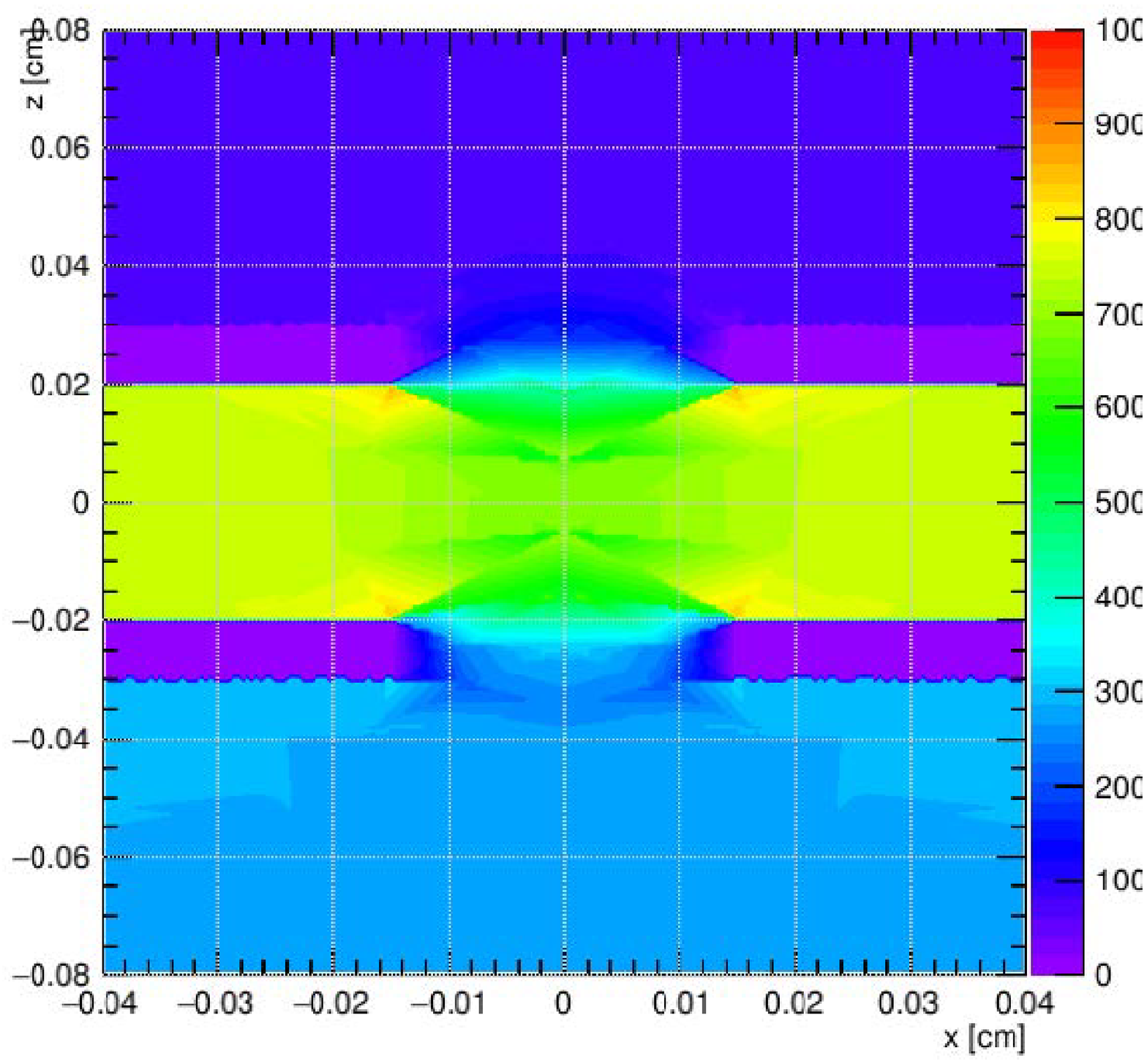}}
\caption{z component of electric field produced in a normal GEM (left) and in a GEM with the same geometry but without dielectric (right).}
\label{simulation}       
\end{figure*}

The first prototype (Fig. \ref{wall_less}) was made of two parallel copper plates 0.2 mm of thickness with a distance between them of 0.4 mm, covered with CuO to make them electrically resistive. These plates were stretched on the supporting frames ensuring a constant distance between them and the holes in the opposite plate were aligned with respect to each other. The spacers were located outside the active area of the detector allowing to considerably increase the length of the dielectric surface between electrodes. In this way, the leak current was reduced to unmeasurable value. Using ANSYS software \cite{lee2018finite}, the electric field maps of a normal GEM and of a wall-less one have been produced. Simulations show that the electric field is quite similar between the two GEM structures: there is a field-line focusing effect along the axis of the aligned holes (Fig. \ref{simulation}).

\begin{figure*}
\centering
\resizebox{0.6\textwidth}{!}{
\includegraphics{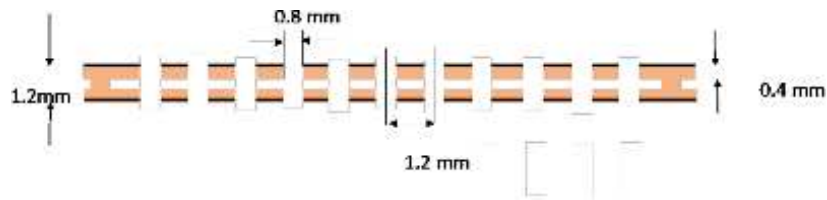}}
\caption{Drawing of the second prototype of $wall-less$ TGEM in which copper oxide electrodes were manufacture on the top of dielectric plates.}
\label{wall_less2}       
\end{figure*}

The second prototype was easier to manufacture, however, its operational voltage at the given gain was higher than in the previous design due to the larger thickness of the avalanche gap. In this prototype the CuO electrode was manufacture on the top of the G-10 plates 0.4 mm thick (Fig. \ref{wall_less2}), supported by 0.4 mm thick spacer located outside the detector active area to reduce the surface leak current. 

\begin{figure*}
\centering
\resizebox{0.6\textwidth}{!}{
\includegraphics{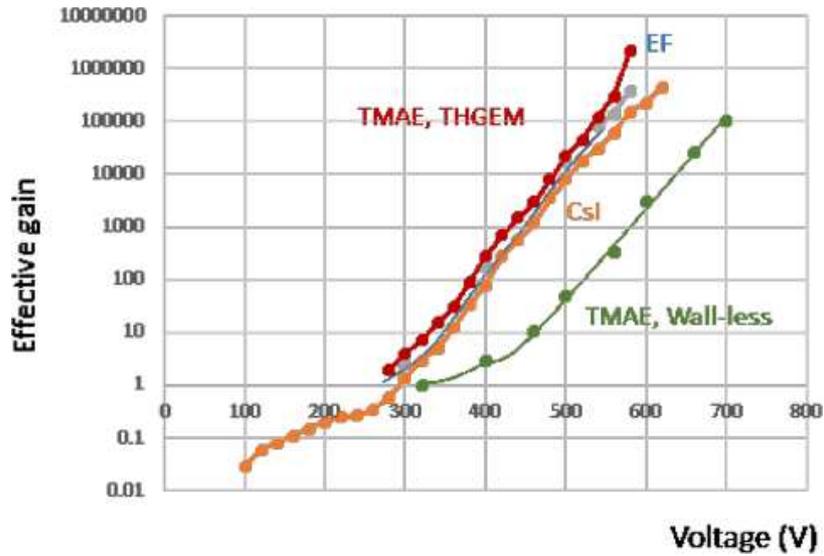}}
\caption{Gas gains vs. applied voltage measured with double TGEMs and wall-less TGEM in mixture of Ne+5\%CO$_2$ (the case of the detector combined  with a CsI photocathode) or in Ne+5\%CO$_2$ with addition of EF or TMAE vapors.}
\label{wallLess_gain}       
\end{figure*}

Both detectors operated perfectly well in gas mixtures with TMAE. In Fig. \ref{wallLess_gain} the gas gain curves for several sealed detectors operated in Ne+5\%CO$_2$ or this mixture with the addition of photosensitive vapors at a total pressure of 1 atm are shown. As can be seen, gains of 10$^5$ or higher were achieved.

\begin{figure*}
\centering
\resizebox{0.7\textwidth}{!}{
\includegraphics{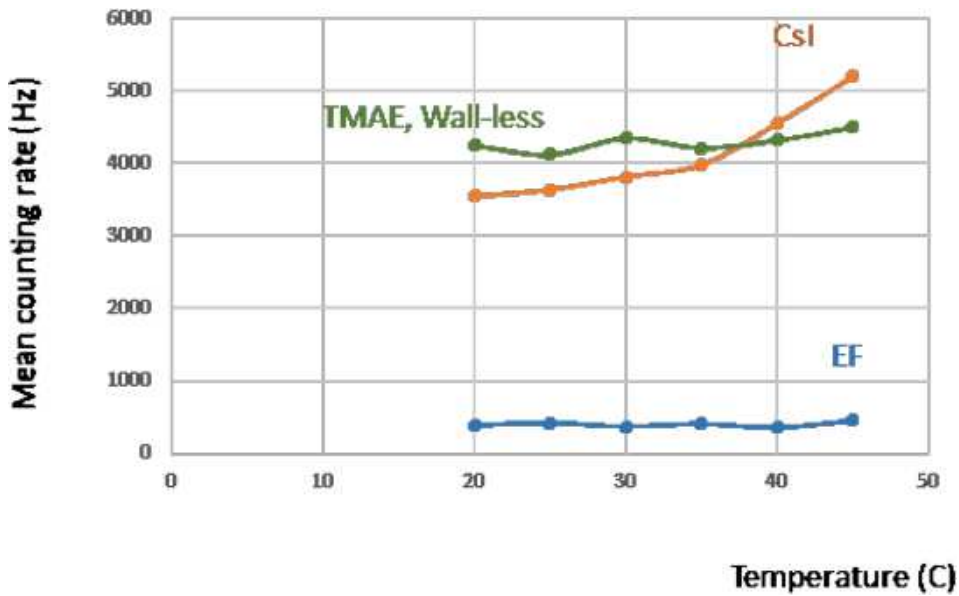}}
\caption{Relative sensitivities of sealed TGEMs and wall-less TGEM as a function of temperature. In these measurements, the overall gain of TGEMs was maintained to be 3$\times$10$^4$.}
\label{sensvstemp}       
\end{figure*}

\begin{figure*}
\centering
\resizebox{0.7\textwidth}{!}{
\includegraphics{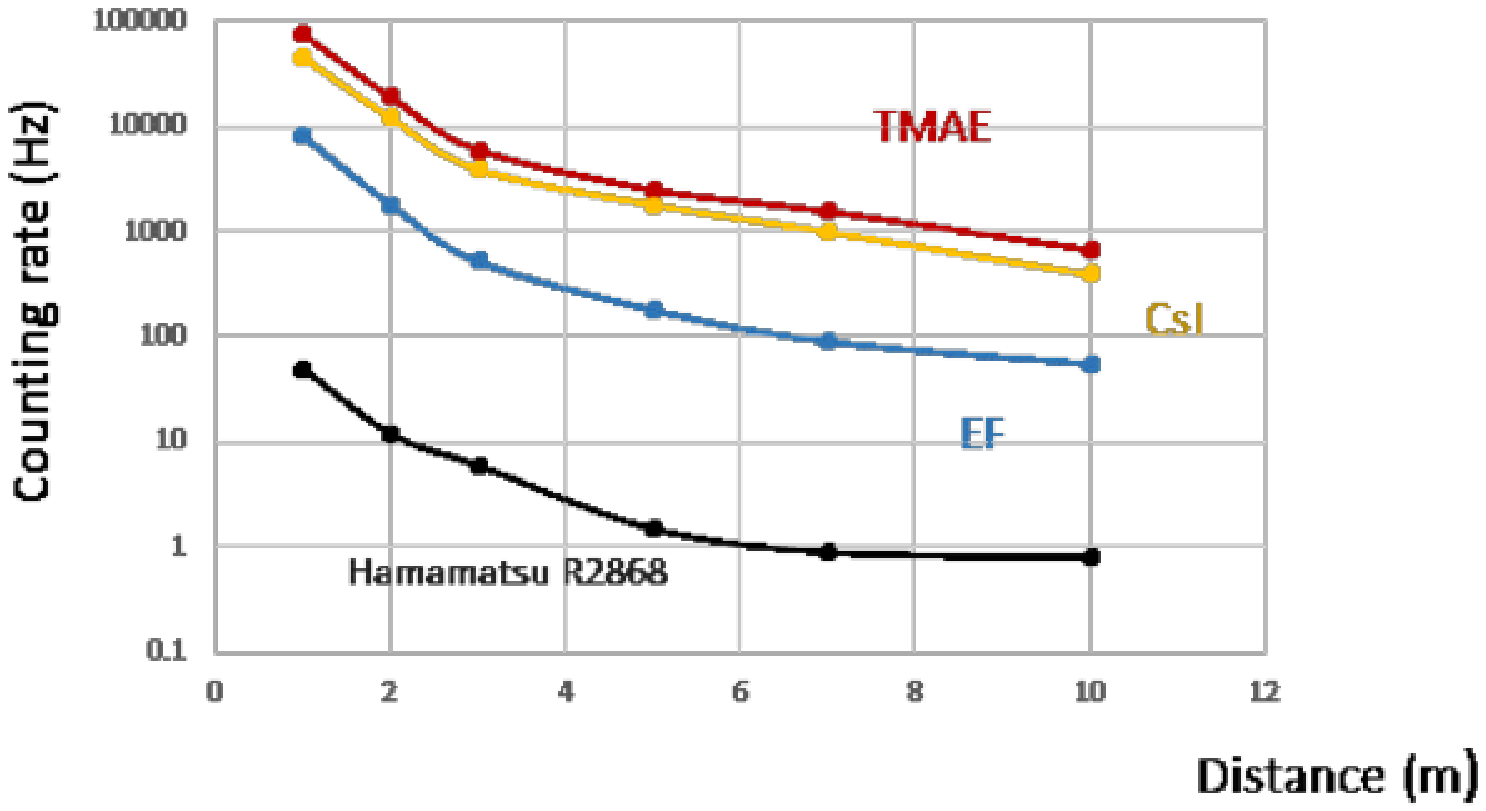}}
\caption{Relative sensitivities of sealed TGEM detectors and Hamamatsu Uvitron R2868 as a function of the distance from the small flame. The gas gain of our detectors was kept to 10$^5$.}
\label{sensvsdist}       
\end{figure*}

In Fig. \ref{sensvstemp} detectors sensitivities (in arbitrary units) vs. temperature are shown. As can be seen, in the case of photosensitive vapors the detectors efficiencies practically do not change with temperature, however, in the case of the CsI photocathode a clear increase was observed. This effect is well known and is connected to the CsI quantum efficiency enhancement with temperature was observed earlier \cite{osti_5232899}, some other solid photocathodes exhibit similar behavior \cite{Buzulutskov:1995kk}.
In Fig. \ref{sensvsdist} a comparison of sealed TGEMs sensitivities with a reference detector, the Hamamatsu R2868, is given. The window size of the gas chamber was practically the same as in the case of single-wire detectors, so it is not astonishing that the relative sensitivities were similar to one presented in Table \ref{table2}. 
The CsI-coated TGEM operated perfectly well inside buildings; however, in direct sunlight condition, it becomes noisy due to the interaction with intense Sun visible radiation. These background pulses were fully suppressed when narrowband filters were used to suppress long-wavelength Sun radiation, in this case, the sensitivity to the flame was reduced 4-6 times (depending on the filter), compared to the data presented in Fig. \ref{sensvsdist}. 

One of the advantages of GEM/TGEM detectors is the possibility of their optimization for the given task. For example, in the case of the reflective CsI photocathode the ratio of the diameter of the holes to their pitch, $d$/$s$, should be slightly decreased compared to the standard value used for the particle detection to enlarge the sensitive area. At the same time, the drift field should be set to be zero or even inverted to improve the photoelectron collections to the holes.
In the case of TMAE vapors, to minimize the undesirable surface photo-effect from the cathode on the top the TGEM, its $d$/$s$ should have the largest value as possible and the drift field should be properly increased and optimized to minimize photoelectron collection from the holes.
Voltages on TGEM electrodes also can be optimized, to meet particular requirements, for example, the reduction of the ion backflow. This could be essential in the case of using reflective or semi-transparent  CsI photocathodes, which quantum efficiency my degrades in time in some gas mixtures  (for example with TMAE) under intense ion bombardment.

\subsection{Output from these studies}

The main advantage of GEM detector is that they are commercially available and has potentials for UV visualization. However, for the efficient single photoelectron detection a multistep configuration is needed which makes the detector more complicated and expensive. Moreover, they are suffering from occasional discharges.
On the other hands, MWPCs have not any discharge problems, and their simplified commercial design with a few wires only (an array of single wire counters), described above could be very attractive.
Thus, is looks that GEM approach is attractive for application when the flame visualization is needed whereas the array of single wire detectors may represent a simple solution. 

\section{Large area incorporated filters: preliminary results}

\begin{figure*}
\centering
\resizebox{0.7\textwidth}{!}{
\includegraphics{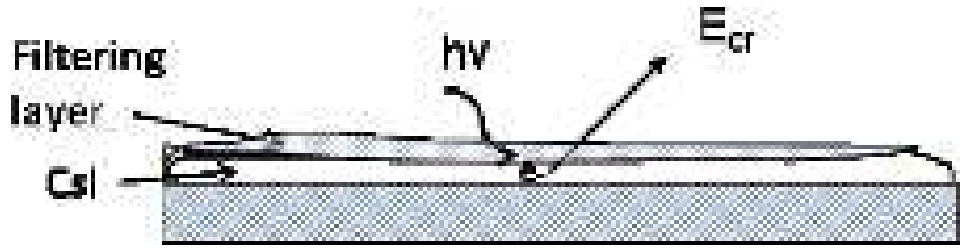}}
\caption{Schematic representation of a CsI photocathode covered with a thin photoelectron filtering layer.}
\label{filtlayer}       
\end{figure*}

\begin{figure*}
\centering
\resizebox{0.6\textwidth}{!}{
\includegraphics{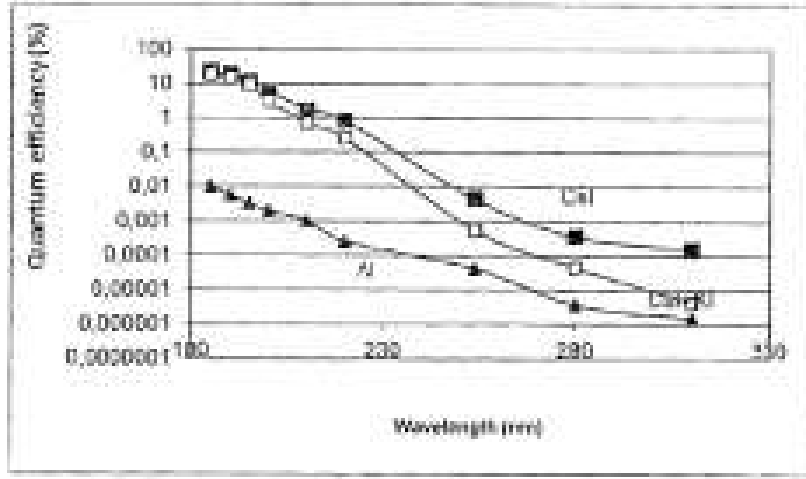}}
\caption{Quantum efficiencies of several photocathodes: bare CsI, CI covered with a KI layer 1nm thick  and a 20 nm (to be checked?) thick Al film.}
\label{qe_pc}       
\end{figure*}

As was already mentioned, one of important advantages of TGEM detectors is its flat-panel geometry allowing to build compact detector with a very large effective area and therefore having the sensitivity on one-two orders of magnitude even higher that was achieved in our present measurements.
Among tested detectors, the most attractive so far is the one with a CsI photocathode due to its high efficiency and excellent temperature stability. However, it requires filters for suppressing long-wavelength radiation and large-are filters are unique and quite expensive.
The pilot tests studies presented below aimed to develop a technology, allowing to suppress the CsI photocathode sensitivity in long wavelengths by coating it surface  with a thin film made from a specially chosen material.
The idea behind this is the following. A photoelectron, created due to the photoelectric effect and escaping the CsI surface, will have a kinetic energy:

\begin{equation}
E_{\rm k} = h\nu - \phi
\label{eq4}
\end{equation}

where $\phi$ is the CsI affinity. If the CsI surface is coated with a film, it will serve as an additional obstacle for the photoelectrons before they can penetrate the gas media of the detector (Fig. \ref{filtlayer}).
For the given film thickness the larger is $E_{\rm k}$, the higher is the probability for the photoelectron to pass through the film. Therefore, by the proper optimization of the film thickness and its martial one can reduce  the photocathode quantum efficiency  in long wavelengths, while preserving its high value in short wavelengths Ref on paten publication
Note, that there were already attempts to cover some solid photocathodes, for example, SbCs with various thin films (CsI or TMAE) \cite{breul}, however, for a completely different purpose: to protect it from the direct contact with the gas which may cause the quantum efficiency  degradation \cite{Charpak:1992xx,Anderson:1993ie,Enomoto:1993bv,BorovickRomanov:1993wk,Peskov:1994ie,Peskov:1995hw,Buzulutskov:1996qc}.
Although in these tests the effect of suppressing photocathode efficiency in long wavelengths was observed but was not exploited for light filtering. However, important conclusions can be driven from the followed studied, performed by several groups \cite{Shefer:1999tr}. One of them is that the most appropriate materials for the covering layer are alkali halides (see figures and some photosensitive liquids, for example, EF, in which transport of free electrons on a reasonably sufficient long distance is possible (Fig. \ref{qe_pc}). In contrast, even relatively thin metallic filters suppress the photocathode efficiency both in long and short wavelengths (Fig. \ref{qe_pc}), although there is some room for their optimization exploiting their $island-type$ structure in small thickness.

Our preliminary results indicate, that a large area photoelectron-filtering layer \cite{peskov_russian} is possible to produce by this technology. Our future task will be to find the best material for this purpose and its optimum thickness. 

\section{Sensors capable to detect simultaneously flames and smoke}

The important advantage of our detectors is that they are operating in proportional mode in which  the signal amplitude is:

\begin{equation}
S_{\rm{d}} = An_{0}
\label{eq5}
\end{equation}

where $n_0$ is the number primary electrons producing the given avalanche and $A$ is a gas gain. The value of $n_0$ obviously depends on a type of radiation: $n_0$ = 1 it in case of  single photoelectrons produced in the detectors by the flame radiation and $n_0$ $>>$ 1 in the case of sparks and cosmic rays. There are several methods allowing to distinguish between sparks and the cosmic radiation, for example to use two flame detectors instead of one and operating in signal coincidence mode, for monitoring the same area. This allows to reliably distinguish between single photons, sparks and cosmic rays. 

\begin{figure*}
\centering
\resizebox{0.7\textwidth}{!}{
\includegraphics{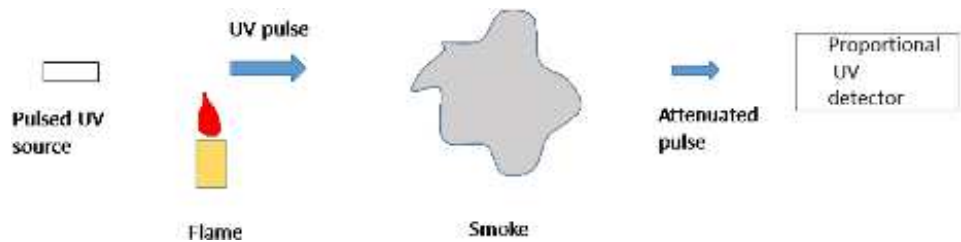}}
\caption{Layout of the flame and smoke detector.}
\label{layout}       
\end{figure*}

Proportional mode offers another unique possibility: to detect simultaneously flames and smokes. Such an apparatus developed by us is schematically show in Fig \ref{layout}.  It contains a smoke detector  described in paragraph \ref{sec:2.1} and a pulse UV source. In preliminary tests a miniature D2 lamp was used producing  periodical UV palsies  with duration of about 10~ns  FWHM. Its emission spectra was typical for any D2 lam: it has a molecular peak at 160 nm and a continuum in the region 200-300~nm. Of course, due to the absorption in the air only radiation with $\lambda$ $>$180 nm was able to reach the detector.

\begin{figure*}
\centering
\resizebox{0.7\textwidth}{!}{
\includegraphics{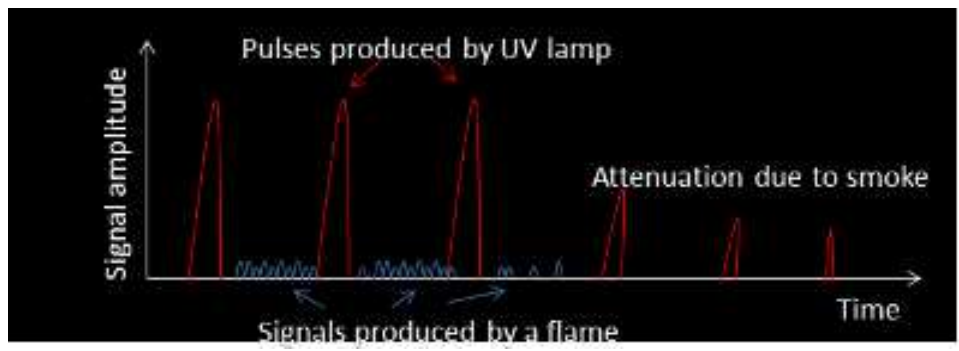}}
\caption{Schematic representation of pulses from a proportional UV  detector, recording at the same time periodic pulses of large amplitudes produced by the pulsed lamp and small amplitude pulses, produced by a flame.}
\label{pulse}       
\end{figure*}

\begin{figure*}
\centering
\resizebox{0.7\textwidth}{!}{
\includegraphics{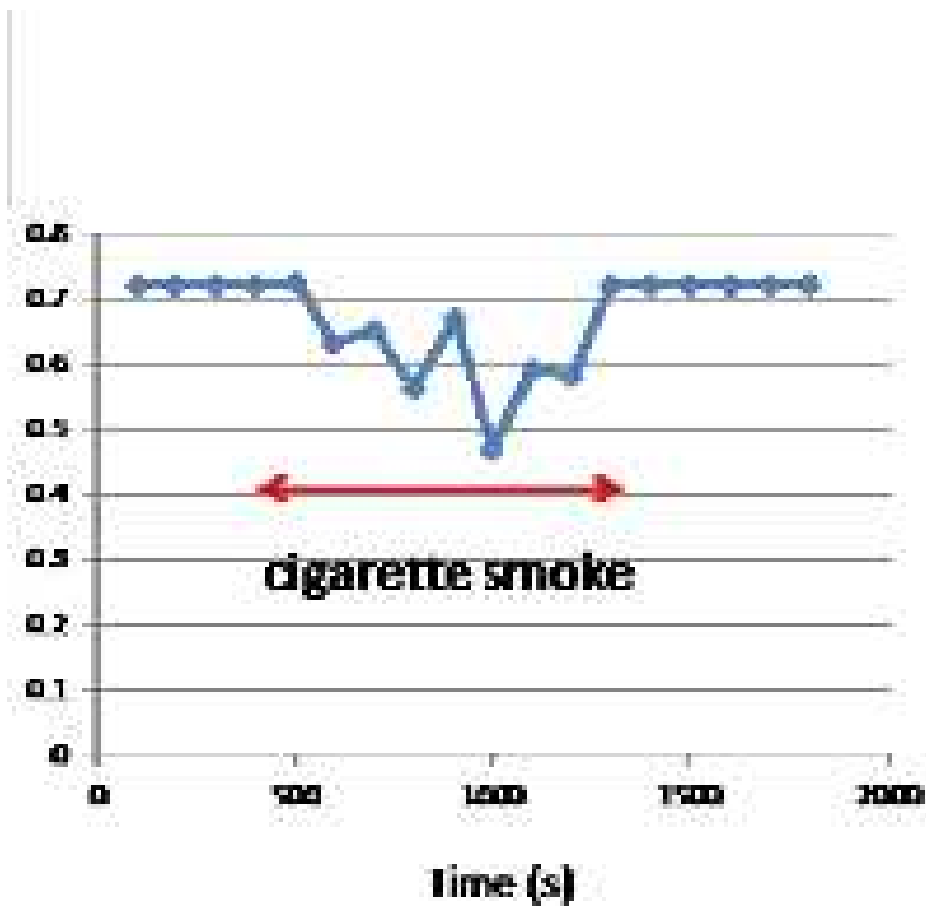}}
\caption{Typical changes in amplitudes of periodical pulses caused by the short term smoke from the cigarettes crossing the pulsed lamp beam.}
\label{cigarette}       
\end{figure*}

The operation principle is illustrated in Fig. \ref{pulse} and Fig. \ref{cigarette}. The UV lamp light generated periodical output pulses of large amplitudes from the detector (due to $n_0$ $>>$ 1). At the same time the detector was able to record pulses with $n_0$ = 1, produced by the flame (if it happened).  If smoke (or any other obstacle) on the way of the UV light it causes its attenuation and the amplitudes of the recorded periodical pulses decreased.

\begin{figure*}
\centering
\resizebox{0.7\textwidth}{!}{
\includegraphics{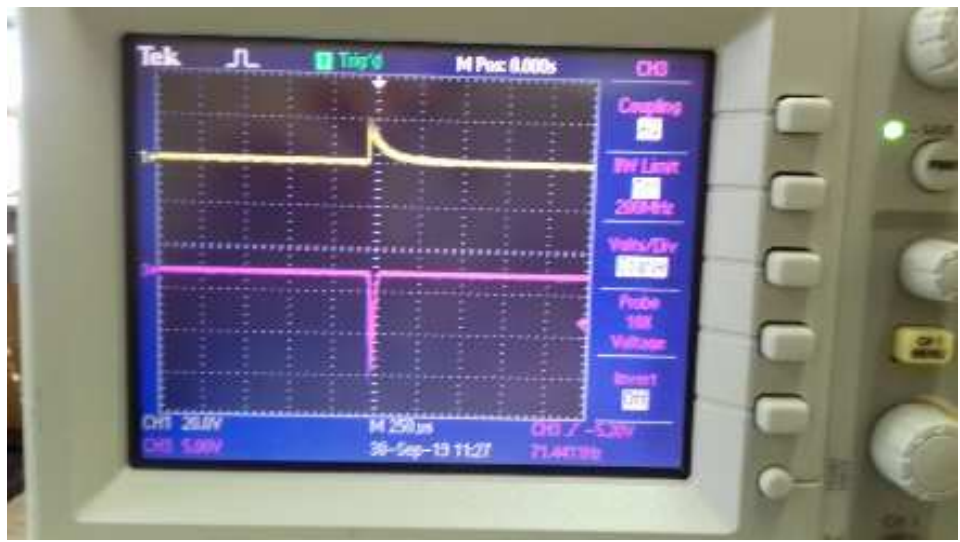}}
\caption{Large amplitude pulses caused by the D2 lamp.}
\label{oscilloscope1}       
\end{figure*}

\begin{figure*}
\centering
\resizebox{0.7\textwidth}{!}{
\includegraphics{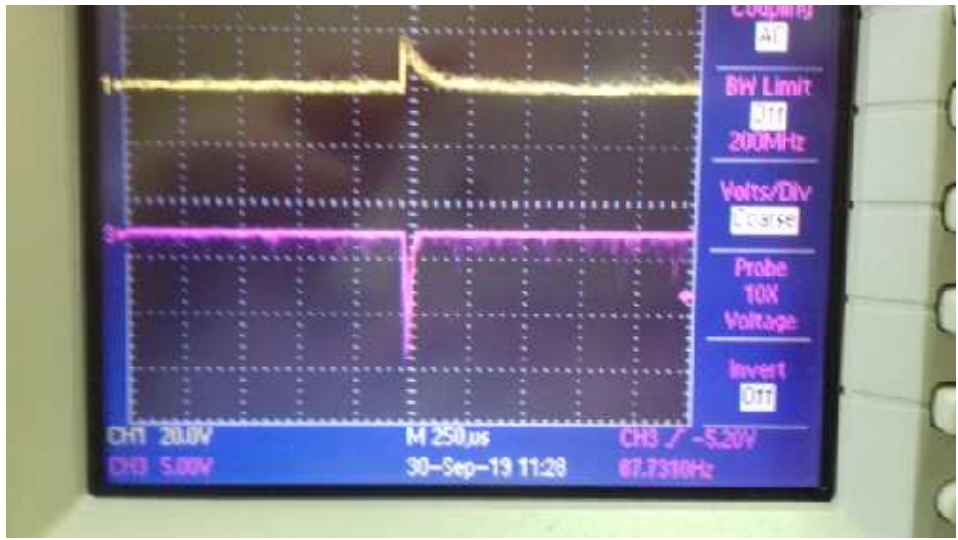}}
\caption{Upper curve: beam-signal from the charge sensitive amplifies. Lower curve: beam signal after the shaper with integration and differential time of 5 ms.}
\label{oscilloscope2}       
\end{figure*}

\begin{figure*}
\centering
\resizebox{0.7\textwidth}{!}{
\includegraphics{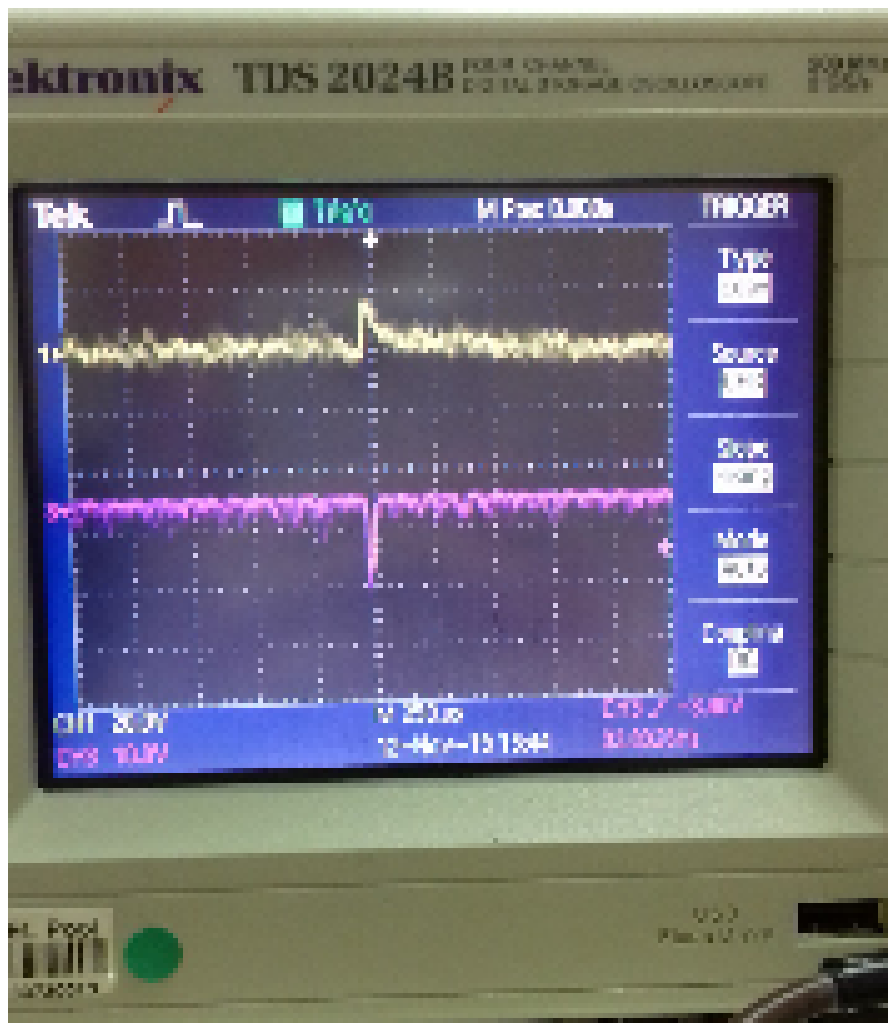}}
\caption{A screen shot from the scope when the detector recorded simultaneous candle and cigarette flame and a smoke produced by the cigarette .}
\label{oscilloscope3}       
\end{figure*}

In Fig. \ref{oscilloscope1},  \ref{oscilloscope2} and  \ref{oscilloscope3} some experimental data are presented. In these measurements the D2 lamp was placed  3~m away for the detector. Although the intensity of the lamp is very low (the sparks ware hardly seen by eyes even at 25~cm and in full darkness), it produced $n_0$ $\approx$ 10 at a distance of 4 m. In Ref. \cite{twiki} one can find video showing some experiments with our flame and smoke detector. 

With much more powerful Xe pulsed lamp (Hamamatsu lamp L7685), combined with a lens, the monitoring distance could be increased on order of magnitudes. When the D2 lamp was  placed in a focal plane of the lense,
the distance was increased up to 10 m.
With such UV lamp, with a light beam splitter with optical fibers, one can monitor a large area of interest at various directions and angles.

\section{Conclusions}

The following commercial prototypes of sealed flame and spark detectors, suitable for some specific  applications, were developed  and extensively tested in this work:

\begin{itemize}
	\item Sealed detectors with CsI photocathodes for indoor applications capable to operate stably in the temperature range [-20$^\circ$C, +45$^\circ$C], with a detection efficiency 1000 times higher than that of the best commercial sensors (Class 1). Combined with   filters, they can be used for outdoor applications. However, the efficiency for flame detection will drop about five times;
        \item solar blind detectors filled with photosensitive vapors: EF and TMAE. These detectors can operate stably in a temperature interval [+20$^\circ$C, +45$^\circ$C];
        \item large-area flat panel detectors with Ni or Cu I photocathodes for temperature range [-70$^\circ$C, +90$^{\circ}$C] suitable for outdoor applications;
        \item finally,  a prototype of an apparatus capable simultaneously  and remotely detect flame, sparks  and smoke were developed.
\end{itemize}


One of the tasks of our studies was to evaluate the possibility to uses GEM-type detectors.
The results obtained proof that GEM-based sealed detectors of flame can be manufactured and they are stable in time. This opens the possibility to use them in commercial flame detectors. Due to the large sensitive area and high quantum efficiency of the photocathode, their sensitivity will be on orders more sensitive than existing commercial sensors. Moreover, in a sophisticated flame monitoring system, one can exploit their imaging capability \cite{Martinengo:2011zz,Peskov:2011ei,Alexeev:2012tba,Volpe:2017eqm}. Combined with pulse UV source they also can be used for smoke and dangerous gases indoor and outdoor survey. We thus think that the described technology is matured and ready for transmission to the industry.

\begin{acknowledgement}
\section*{Acknowledgements}

These studies were performed in the framework of the ATTRACT-SMART project.
For the technical support we thank M. Meli, J. Ven Belleviin and M. Van Stenis.

\end{acknowledgement}

%

%
%

%
 \bibliographystyle{utphys}
 \bibliography{smart}

%
%
%

\end{document}